\documentclass[runningheads,natbib]{svjour2}
\bibpunct{[}{]}{;}{n}{}{,} 
\smartqed  
\usepackage{graphicx}  
\usepackage{mathptmx}      
\newcommand{\chandra}{\textit{Chandra}}
\newcommand{\xmm}{\textit{XMM-Newton}}
\newcommand{\einstein}{\textit{Einstein}}
\newcommand{\rosat}{\textit{ROSAT}}
\newcommand{\hst}{\textit{HST}}
\newcommand{\vla}{\textit{VLA}}
\newcommand{\evla}{\textit{EVLA}}
\newcommand{\asca}{\textit{ASCA}}
\newcommand{\atca}{\textit{ATCA}}
\newcommand{\alma}{\textit{ALMA}}
\newcommand{\spitzer}{\textit{Spitzer}}
\newcommand{\herschel}{\textit{Herschel}}
\newcommand{\emerlin}{\textit{e-MERLIN}}
\newcommand{\suzaku}{\textit{Suzaku}}

\journalname{Astronomy and Astrophysics Review}
\begin{document}

\title{The X-ray Jets of Active Galaxies}
\author{D.M. Worrall}
\institute{HH Wills Physics Laboratory, University of Bristol, Tyndall Ave,
Bristol BS8 1TL, UK \\
\email{d.worrall@bristol.ac.uk}}
\date{The Astronomy and Astrophysics Review, Volume 17, 1-46 (2009),
  Fig 5 updated}
\maketitle

\begin{abstract}
Jet physics is again flourishing as a result of \chandra's ability to
resolve high-energy emission from the radio-emitting structures of
active galaxies and separate it from the X-ray-emitting thermal
environments of the jets.  These enhanced capabilities have coincided
with an increasing interest in the link between the growth of
super-massive black holes and galaxies, and an appreciation of the
likely importance of jets in feedback processes.  I review the
progress that has been made using \chandra\ and \xmm\ observations of
jets and the medium in which they propagate, addressing several
important questions, including:   Are the radio structures in a state of minimum
energy?  Do powerful large-scale jets have fast spinal speeds?  What
keeps jets collimated? Where and how does
particle acceleration occur? What is jet plasma made of? 
What does X-ray emission tell us about
the dynamics and energetics of radio plasma/gas interactions?
Is a jet's fate determined by the
central engine?
\end{abstract}

\tableofcontents

\section{The stage is set}
\label{sec:stage}

\subsection{Historical perspective}
\label{sec:historical}

In the 1970s and 1980s the powerful capabilities of radio
interferometry gave birth to the study of extragalactic radio jets.
It became clear that radio jets are plasma outflows originating in the
centres of active galaxies, seen through their synchrotron emission.
After much debate, properties such as the relative one-sidedness of
the jets, and the measurement of apparent superluminal expansion, by Very
Long Baseline Interferometry (VLBI), were accepted as due to the
outflows having relativistic bulk speeds.  Early attempts at unifying
source populations based on special relativity and apparent source
properties \citep[e.g.,][]{scheuer-readhead} have developed over the
years into comprehensive unified schemes \citep[e.g.,][]{barthel}
whereby quasars are explained as radio galaxies whose jets are at
small angles to the line of sight and so are boosted by relativistic
effects.

By the mid 1990s, the study of radio jets had reached something of a
hiatus, and major groups around the world turned their attention to
other pursuits such as gravitational lensing and the study of the Cosmic
Microwave Background (CMB) radiation.  A turning point was the
sensitivity and high-fidelity mirrors of the \chandra\ X-ray
Observatory \cite{weisskopf}, which resulted in the detection of
resolved X-ray emission from many tens of well-known extragalactic
radio sources (see \cite{harris-kraw} for a source compilation as of
2006: the number continues to increase).  When combined with X-ray 
measurements of the
ambient gas made with \chandra\ and \xmm, and multiwavelength data,
many important questions related to the physics of jets can be
addressed.  Progress towards answering those questions is the
substance of this review.

The enhanced capabilities for the X-ray study of jets have coincided
with strong interest from the wider astronomical community in the
growth of supermassive black holes (SMBHs), following the links that
have been made between SMBH and galaxy growth
\citep[e.g.,][]{richstone, gebhardt}.  SMBHs (and indeed compact objects
of stellar mass) commonly produce jets, as an outcome of accretion
processes responsible also for black-hole growth.  It is also clear
that extragalactic jets are capable of transferring large amounts of
energy to baryonic matter in the host galaxies and surrounding
clusters at large distances from the SMBH.  The way in which heating
during the jet mode of AGN activity might overcome the problem of fast
radiative cooling in the centre of clusters is now intensely studied in
nearby objects (see \S \ref{sec:dynamics}), and heating from `radio
mode' activity is included in simulations of
hierarchical structure formation \citep[e.g.,][]{croton}.  We need
therefore to understand what regulates the production of jets and how
much energy they carry.  X-ray measurements of nuclear emission
probe the fueling and accretion processes, and those of resolved jet
emission and the surrounding gaseous medium probe jet composition,
speed, dynamical processes, energy deposition, and feedback.

\subsection{Radiation processes}
\label{sec:processes}

The two main jet radiation processes are synchrotron radiation and
inverse-Compton scattering. Their relative importance depends
on observing frequency, location within the jet, and the speed of the jet.
The thermally X-ray-emitting medium into which the jets propagate
plays a major r\^ole in the properties of the flow and the appearance
of the jets.  The physics of the relevant radiation processes are well
described in published work \citep[e.g.,][]{ginzburg, blum-gould,
pachol, spitzer, rlightman, sarazin, longair}, and most key equations
for the topics in this review, in a form that is independent of the
system of units, can be found in \cite{worr-santiago}.

It is particularly in the X-ray band that synchrotron radiation
and inverse-Compton emission are both important.  X-ray synchrotron
emission depends on the number of high-energy electrons and the
strength and filling factor of the magnetic field in the rest frame of
the jet. Inverse Compton X-ray emission depends on the number of
low-energy electrons, the strength of an appropriate population of
seed photons (such as the CMB, low-energy jet synchrotron radiation,
or emission from the central engine), and the geometry of scattering
in the rest frame of the jet.  In an ideal world, observations would
be sufficient to determine the emission process, and this in turn
would lead to measurements of physical parameters.  In reality, X-ray
imaging spectroscopy, even accompanied by good measurements of the
multiwavelength spectral energy distribution (SED), often leaves
ambiguities in the dominant emission process.  Knowledge is
furthered through intensive study of individual sources or
source populations.

\subsection{Generic classes of jets}
\label{sec:classes}

In discussing jets, it is useful to refer to the Fanaroff and Riley
\cite{fr} classification that divides radio sources broadly into two
morphological types, FRI and FRII.  A relatively sharp division
between FRIs and FRIIs has been seen when sources are mapped onto
a plane of radio luminosity and galaxy optical luminosity
\cite{ledlow-owen} -- the so called Ledlow-Owen relation. FRIIs
are of higher radio luminosity, with the separation between the
classes moving to larger radio luminosity in galaxies that are
optically more massive and luminous. The
distinct morphologies \citep[e.g.,][]{muxlow} are believed to be a
reflection of different flow dynamics \citep[e.g.,][]{leahy}.

\begin{figure}[t]
\centering
\includegraphics[height=5.0cm]{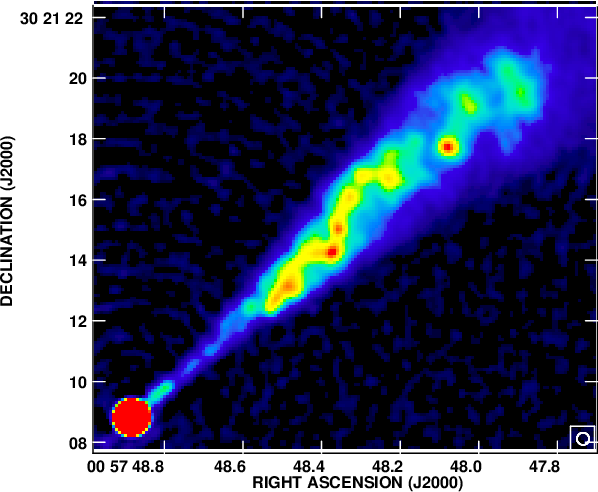}
\includegraphics[height=5.0cm]{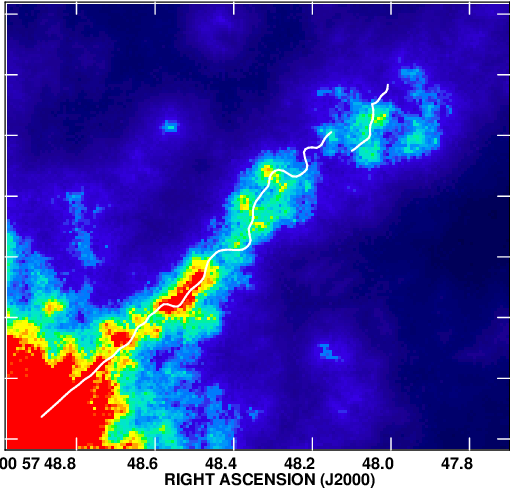}
\caption{Roughly 6.6~kpc (projected)
of the inner jet of the $z=0.0165$ FRI radio galaxy NGC~315.
Left: 5~GHz \vla\ radio map showing a knotty filamentary structure in
diffuse emission. Right: Smoothed \chandra\ X-ray image
of $\sim 52.3$~ks livetime
also showing knotty structure embedded in diffuse emission.
The ridge-line defined by the radio structure is shown in white,
and indicates a level of correspondence between the
radio and X-ray knots. Figure adapted from \protect\cite{worr-N315}.
}
\label{fig:n315jet}      
\end{figure}

FRI sources (of lower isotropic radio power, with BL Lac
objects as the beamed counterpart in unified schemes) have
broadening jets feeding diffuse lobes or plumes that can show
significant gradual bending, usually thought to be due to ram-pressure as the
source moves relative to the external medium.  The jet emission is of high
contrast against diffuse radio structures, implying that the jet
plasma is an efficient radiator. kpc-scale jets are usually
brightest at a flaring point some distance from the 
active galactic nucleus, and then fade
gradually in brightness at
larger distances from the core, although this pattern is often
interrupted by bright knots seen when the jet is viewed in the radio
or the X-ray.  Such an example is shown in Figure~\ref{fig:n315jet}\footnote{Values for the
 cosmological parameters of $H_0
= 70$~km s$^{-1}$ Mpc$^{-1}$, $\Omega_{\rm {m0}} = 0.3$, and
$\Omega_{\Lambda 0} = 0.7$ are adopted throughout this review.}.
The jets are believed
to slow from highly-relativistic to sub-relativistic flow on
kpc-scales from entrainment of the external interstellar medium (ISM),
perhaps enhanced by stellar mass loss within the jet.  The strong
velocity shear between the jet flow and the almost stationary external
medium must generate instabilities at the interface
\cite{birkinshaw-kh}, and drive the flow into a turbulent state.  The
physics of the resulting flow is far from clear, although it can be
investigated with simplifying assumptions \citep[e.g.,][and see \S
\ref{sec:jetcollimate}]{bick-31, bick-315}.

\begin{figure}[t]
\centering
\includegraphics[height=6.0cm]{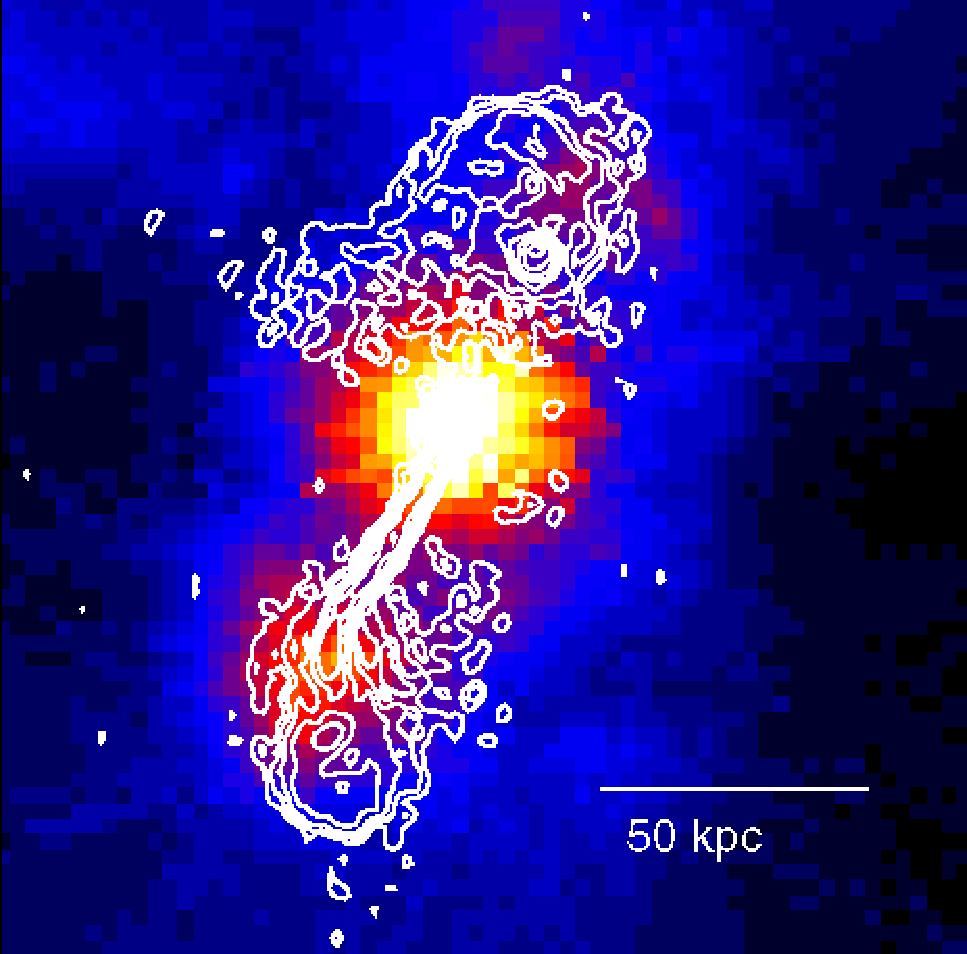}
\caption{The $z=0.458$ FRII radio galaxy 3C~200.
A smoothed 0.3-5~keV \chandra\ X-ray image of $\sim 14.7$~ks livetime  is
shown with radio contours from a 4.86~GHz \vla\ radio map \cite{leahy-atlas}
(beam size $0.33''\times 0.33''$).  Both nuclear \protect\cite{belsole-cores}
and extended X-ray emission are detected.
A rough correspondence of some of the extended X-ray emission with the
radio lobes has resulted in the claim for inverse-Compton scattering
of the CMB by electrons in the lobes \protect\cite{croston-lobes}, but most of the
extended emission over larger scales is now attributed
to cluster gas \protect\cite{belsole-env}.
}
\label{fig:3c200}      
\end{figure}

FRII sources (of higher isotropic radio power, with quasars as the
beamed counterpart in unified schemes) have narrower jets that are
sometimes faint with respect to surrounding lobe plasma and that
terminate at bright hotspots (Fig.~\ref{fig:3c200}).  The jets are
often knotty when observed with high resolution, and the jets can bend
abruptly without losing significant collimation (see \S\ref{sec:0637}
and \S\ref{sec:FRIpolariz} for examples). The bending is often
large in quasar jets, supporting the conjecture that quasars are
viewed at small angle to the line of sight and that bends are
amplified through projection.  In contrast to FRI jets which are in
contact with the external medium, the standard model for FRII jets is
that they are light, embedded in lobe plasma, and remain supersonic
with respect to the external gas out to the hotspots.  The energy and
momentum fluxes in the flow are normally expected to be sufficient to
drive a bow shock into the ambient medium.  The ambient gas, 
heated as it crosses the shock, forces old jet material that has
passed through the hotspots into
edge-brightened cocoons.  FRII jets are thus low-efficiency radiators
but efficient conveyors of energy to large distances.  They are often
hundreds of kpc in length (particularly when deprojected for their
angles to the line of sight), crossing many
scale heights of the external medium from relatively dense gas in a
galaxy core to outer group or cluster regions where the external
density and pressure are orders of magnitude lower.  State-of-the-art
three-dimensional magneto-hydrodynamical simulations that incorporate
particle transport and shock acceleration do well at reproducing the
essential characteristics of synchrotron emission from such a source,
and suggest that the shock and magnetic-field structures of the
hotspots and lobes are extraordinarily complex and unsteady
\cite{tregillis, tregillis2}.

\subsection{Lifetimes and duty cycles}
\label{sec:dutycycles}

Individual FRI and FRII radio galaxies are thought to live for at most
some tens of millions of years \citep[e.g.,][]{mack, kaiser}.  Age
estimates are based on measuring curvature in the radio spectra caused
by radiative energy losses of the higher-energy electrons over the
lifetime of the sources \citep[e.g.,][]{alexanderleahy}.  In contrast
to the relative youth of observed radio structures, present-day
clusters were already forming in the young Universe.  Ideas that radio
sources have an important r\^ole in heating cluster gas (see
\S\ref{sec:dynamics}) then require a correct balance between the
duty-cycle of repeated radio activity and heating efficiency as a
function of jet luminosity.  The duty cycle can be probed by searching
for evidence of repeated activity from individual sources.  Radio
sources classified as GHz-Peaked Spectrum (GPS) or Compact Steep
Spectrum (CSS) are small and believed to be either young or have their
growth stunted by the external medium \cite{odea-GPS}, and source
statistics suggest that if they evolve to kpc-scale sizes they must
dim while so doing \cite{readhead}.  VLBI kinematic studies provide
convincing evidence that sources in the Compact Symmetric Object (CSO)
subset, at least, are young, with current ages less than $10^4$ years
\cite{conway}.  The fact that it is relatively uncommon to see GPS
sources with extended radio emission that may be a relic of previous
activity has been used to argue that periods between
sustained activity are generally at least ten times longer than the
radiative lifetime of the radio emission from the earlier activity
\cite{stanghellini}.  This is consistent with a time between
episodes of activity in
FRIIs of between about $5 \times 10^8$ and
$10^9$ years that is estimated using optical- and radio-catalog cross correlations
coupled with an average source lifetime of about $1.5 \times 10^7$ years from
modelling projected source lengths \cite{bird}.
Of course, within the lifetime of an individual radio source there
might be shorter-term interruptions or variations of activity 
(see \S\ref{sec:evolution}).

\section{Are the radio structures in a state of minimum energy?}
\label{sec:minenergy}

\subsection{Calculation of the minimum-energy field}
\label{sec:minenergycalc}

The magnetic field strength and particle spectrum
are important for jet physics as they define the internal pressure.
The level of
synchrotron radiation depends on the magnetic-field strength and the
number of relativistic electrons and positrons, but these quantities are
inseparable based on the observed synchrotron radiation alone.
To progress further it is usual to assume that the
source is radiating such that its combined energy in relativistic particles and
magnetic field is a minimum \cite{burbidge}.  In this situation the
energy in the magnetic field is $\sim 3/4$ of the energy in the
relativistic particles, and so this is similar to the condition in which the two
are equal and the source is in `equipartition'.  A change in any
direction of the ratio of energy density in particles to magnetic
field increases the total energy and pressure in the emitting plasma.

The minimum-energy magnetic field for a power-law spectrum of
electrons producing radiation of a measured flux density at a
particular frequency can be calculated analytically
\citep[e.g.,][]{worr-santiago}, and for more complicated spectra the
results can be obtained via numerical integration.  
Physical insight can be gained by considering a
power-law spectrum where electrons give rise to
a synchrotron luminosity, $L_\nu$, at a given frequency $\nu$ of the form

\begin{equation}
L_\nu \propto \nu^{-\alpha}.
\label{eq:freqspectrum}
\end{equation}

\noindent
It is now normally
thought preferable to define the spectral limits via a minimum and
maximum Lorentz factor for the electrons in the source frame,
$\gamma_{\rm min}$ and $\gamma_{\rm max}$,
\citep[e.g.,][]{worr-santiago}, rather than as synchrotron frequencies
in the observer's frame \citep[e.g.,][]{miley}, since the former is
related to acceleration processes and has the potential for being
chosen on a physical basis.  
Except in the special case of $\alpha = 0.5$, 
the minimum-energy magnetic field strength, $B_{\rm me}$, is given by

\begin{equation}
B_{\rm me} = \left[{(\alpha + 1)  C_1 \over 2 C_2} {(1 + K)\over \eta
V} L_\nu
\nu^{\alpha}
 {\left(
\gamma_{\rm max}^{1 - 2\alpha} - \gamma_{\rm min}^{1 -
2\alpha}
\right) \over (1 - 2\alpha)}\right]^{1/(\alpha + 3)} ,
\label{eq:we6}
\end{equation}

\noindent
where $V$ is the source volume, and
$C_1$ and $C_2$ are combinations of fundamental physical constants and
functions of $\alpha$ given by synchrotron theory \citep[for
details see][]{worr-santiago}.
Following the notation of \cite{miley}, 
$K$ is the ratio of energy in other relativistic
particles to that in the electron and positron component,
and $\eta$ is the fraction of the
volume filled by particles and fields (the so-called filling factor).
The true minimum energy is when the only relativistic particles are
radiating leptons, and the volume is completely and uniformly
filled with radiating particles and fields. Some authors
consistently use these assumptions when calculating $B_{\rm me}$.  
If $K > 0$ or $\eta < 1$ then $B_{\rm me}$ is increased. 
Results for $B_{\rm me}$ are more strongly dependent on
$\gamma_{\rm min}$ than $\gamma_{\rm max}$, since $\alpha > 0.5$
for most observed radio spectra.

Relativistic beaming of a source affects $B_{\rm me}$ (as considered
later in \S \ref{sec:iC-CMB}).  Since there is
inevitably uncertainty in the value of beaming parameters,
$B_{\rm me}$ is best measured in components for which
bulk relativistic motion is believed to be small or negligible.
Of course, even in the absence of relativistic beaming, the angle to the line of
sight, $\theta$, enters into the calculation via a correction from
projected linear size into true source volume, $V$.
Typical values found for $B_{\rm me}$ in radio lobes and hotspots are
2--200 $\mu$Gauss (0.2--20 nT) \citep[e.g.,][]{kataokasta}, although a
hotspot field as large as 3000 $\mu$Gauss has been measured
\cite{godfrey}.

\begin{figure}[t]
\centering
\includegraphics[height=5.75cm]{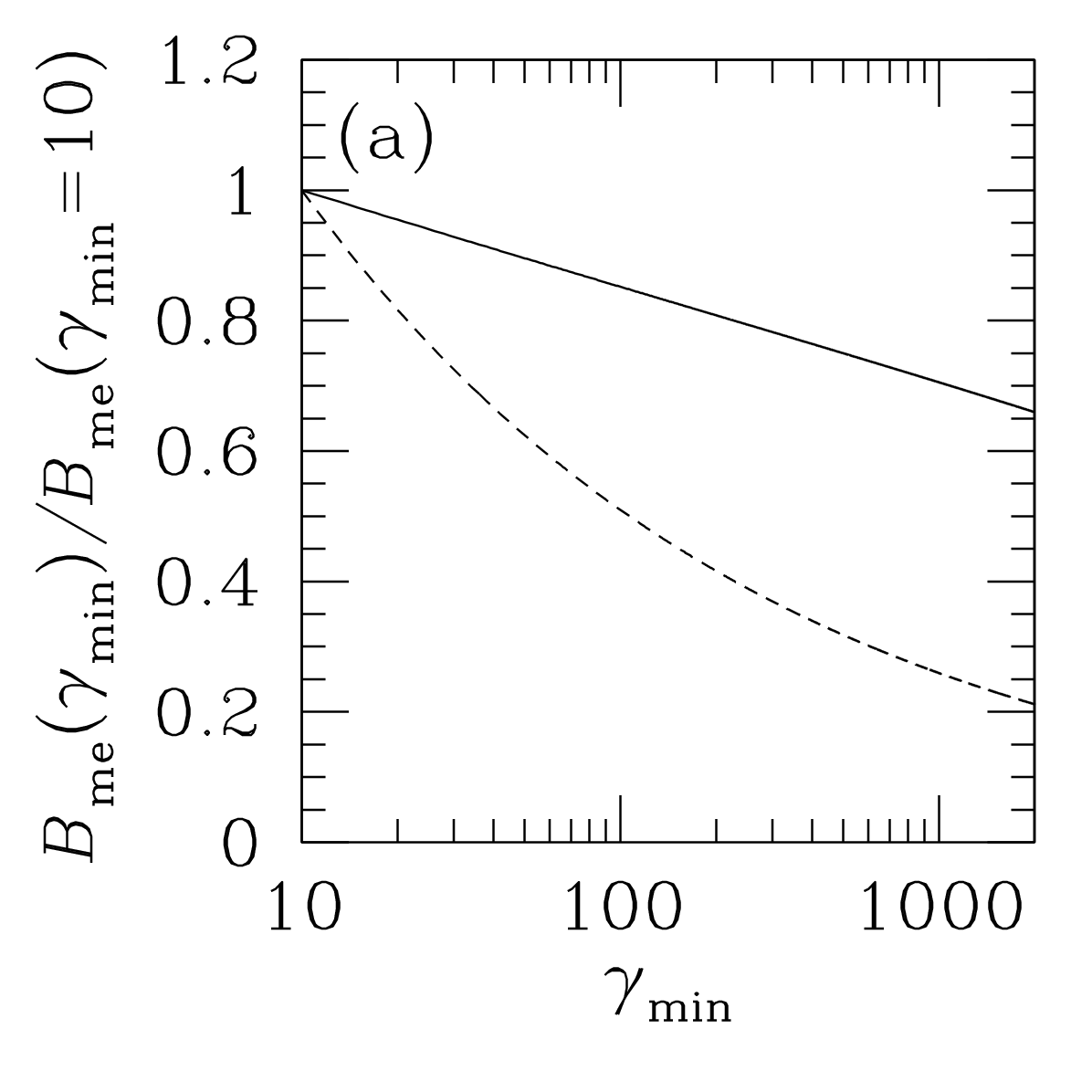}
\includegraphics[height=5.75cm]{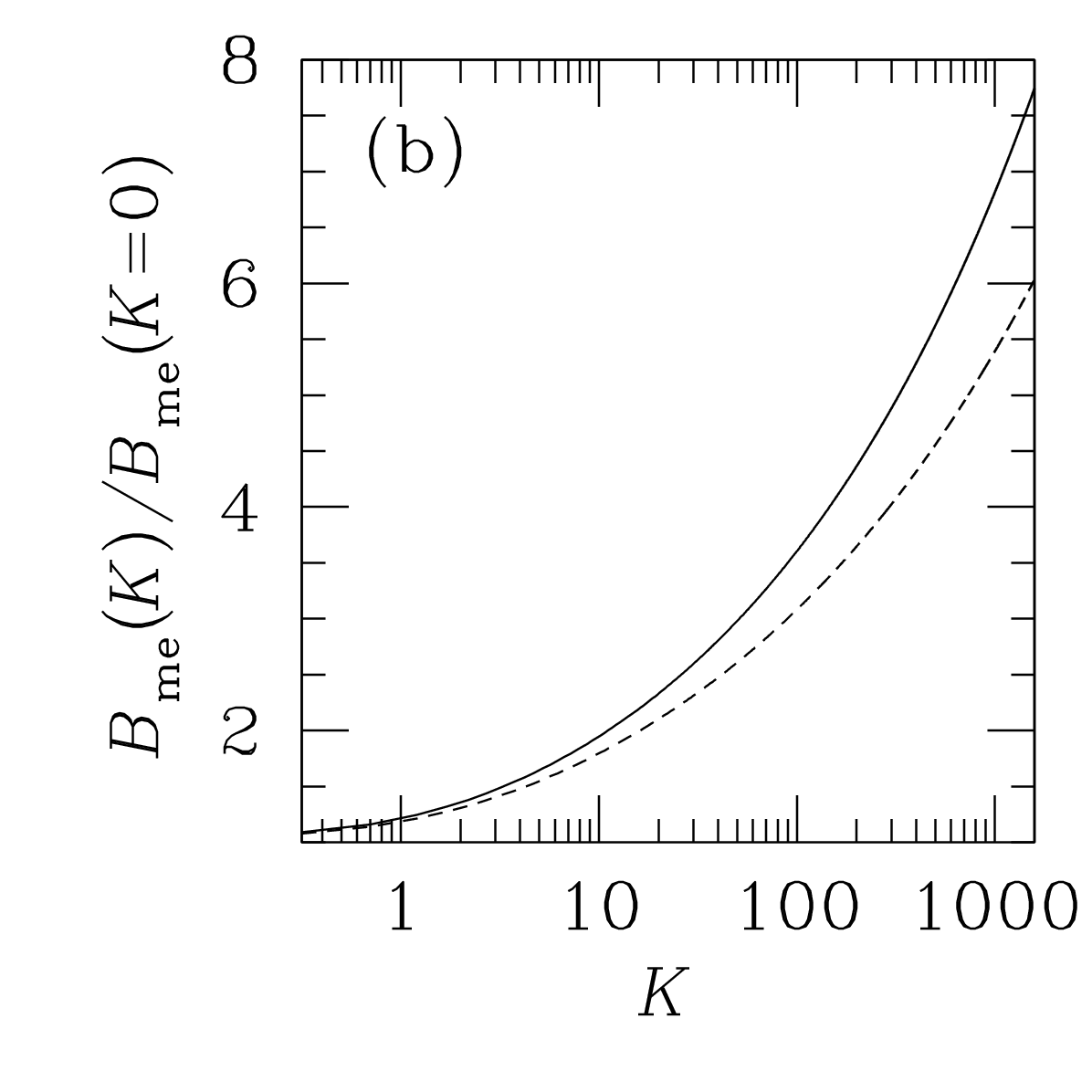}
\includegraphics[height=5.75cm]{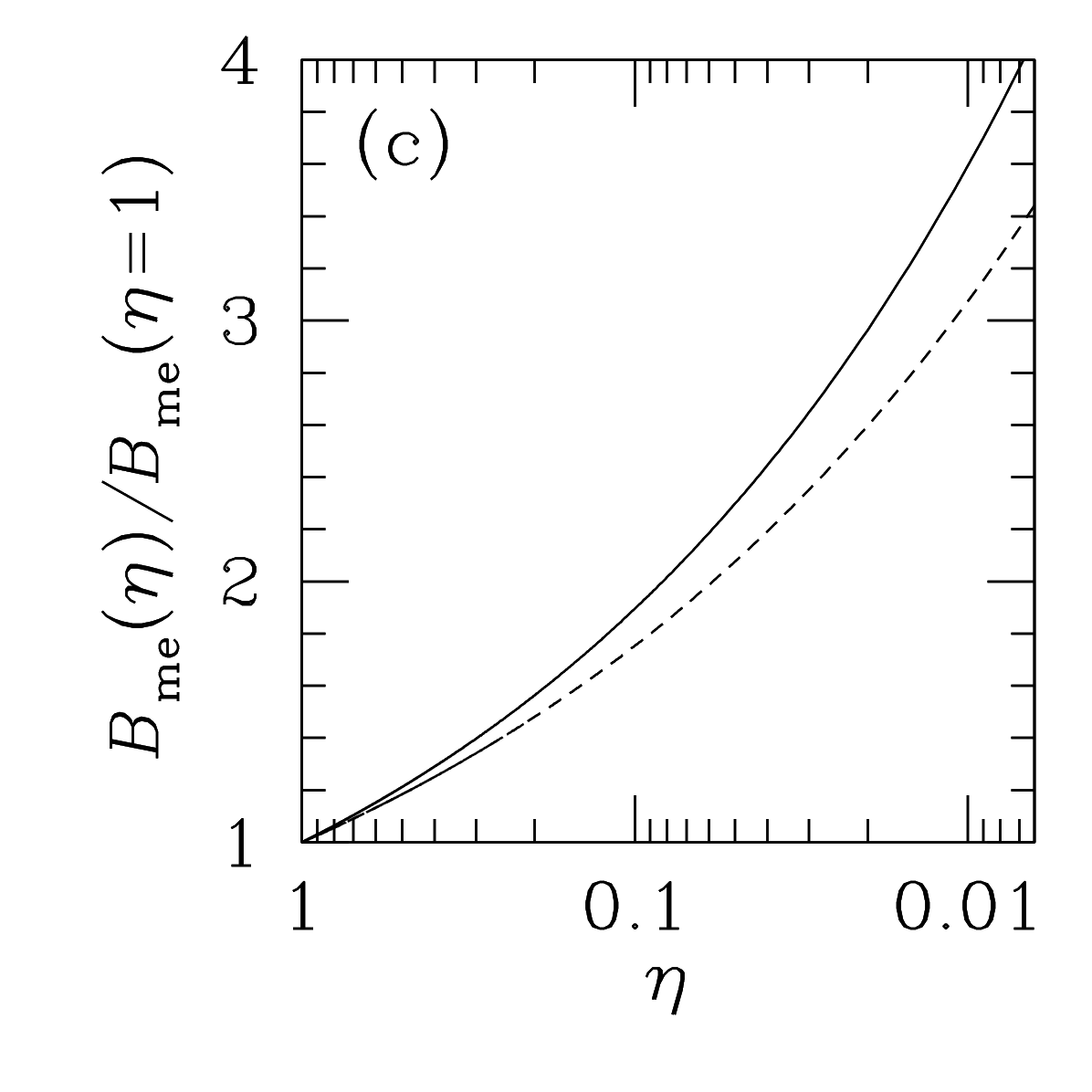}
\includegraphics[height=5.75cm]{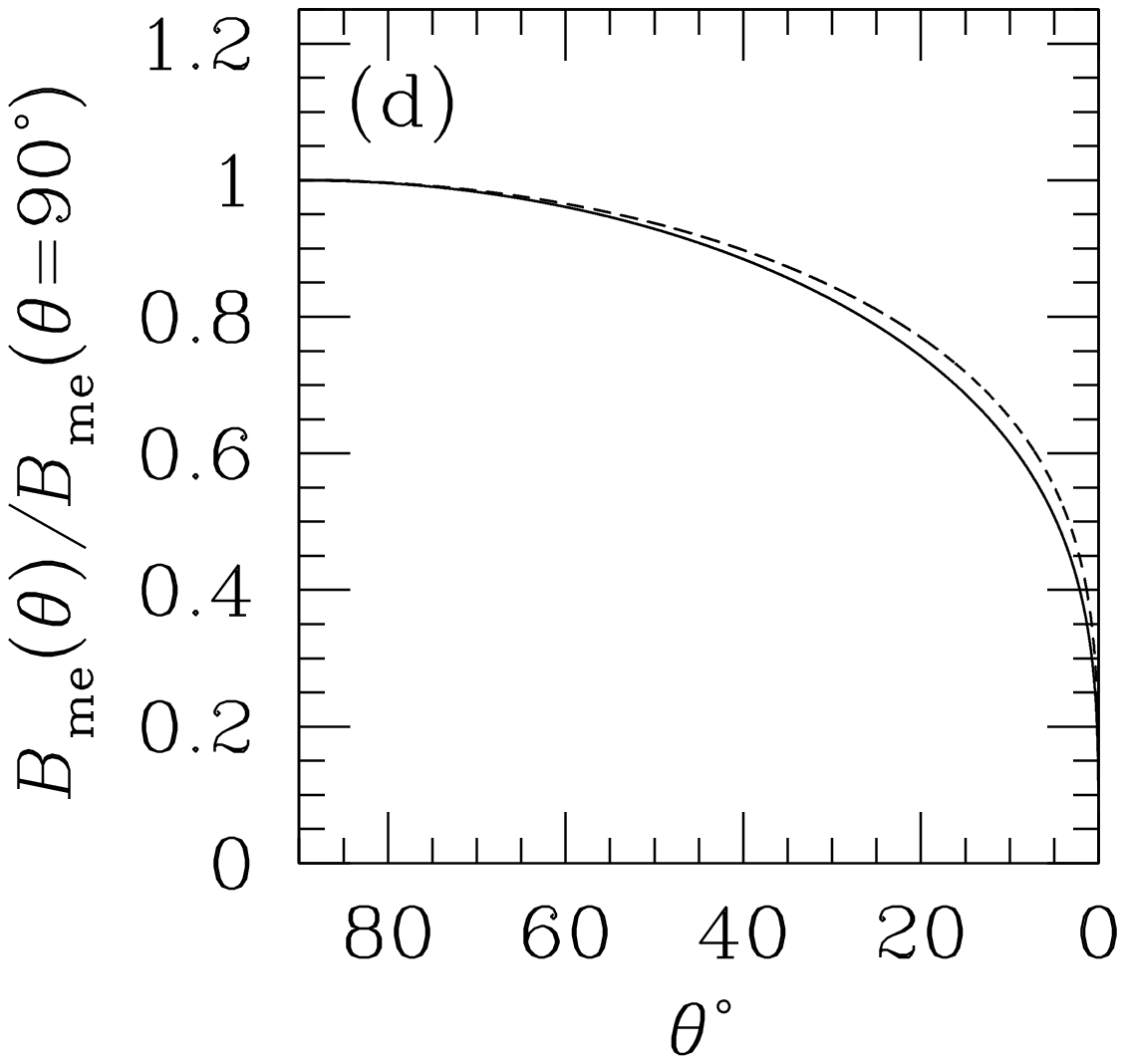}
\caption{Effect on the calculated minimum-energy magnetic field
if a parameter value is varied from its nominal value (left-hand-side
of plot).
Results are for a power-law electron spectrum, extending from
Lorentz factor $\gamma_{\rm min}$ to $\gamma_{\rm max} = 10^5$, that
gives rise to a synchrotron spectrum $S_\nu \propto \nu^{-\alpha}$
with $\alpha=0.6$ (solid lines) and $\alpha=1.1$  (dashed lines).
(a): increasing $\gamma_{\rm min}$ from a value of 10.  (b):
increasing the ratio of
energy in other particles to that in electrons, $K$, from a value of
zero. (c): decreasing
the filling factor, $\eta$, from a value of 1. (d) decreasing the
angle to the line of sight, and thus increasing the source volume
from the projected size at $\theta =90^\circ$.}
\label{fig:bme}      
\end{figure}

Figure~\ref{fig:bme} shows the dependence of $B_{\rm me}$ on
$\gamma_{\rm min}$, $K$, $\eta$, and $\theta$,
separately for electrons giving rise to synchrotron spectra
with $\alpha = 0.6$ and $\alpha
= 1.1$.  The former slope is as expected from electrons
undergoing highly relativistic shock acceleration \cite{achterberg},
and the latter where energy losses have steepened the spectrum.
The curves show that $B_{\rm me}$ changes rather little (within
factors of at most a few) for rather large changes in the input
assumptions.

\subsection{Using X-rays to test minimum energy}
\label{sec:minenergytest}

The minimum-energy assumption can be tested by combining measurements
of synchrotron and inverse-Compton emission from the same electron
population.  If the inverse Compton process is responsible for most of
the X-ray radiation that is measured, and the properties of the photon
field are known, the X-ray flux density is proportional merely to the
normalization of the electron spectrum, $\kappa$, if the usual
power-law form

\begin{equation}
N_{\rm e}^{\rm (rel)} = \kappa \gamma^{-p} \quad (\gamma_{\rm min} \leq
\gamma \leq \gamma_{\rm max}) 
\label{eq:electronspectrum}
\end{equation}

\noindent
is assumed, where $N_{\rm e}^{\rm (rel)}$ is the number of
relativistic electrons per unit $\gamma$.  The upscattered photons
might be the CMB, whose properties are well known.  Alternatively they
could be the radio synchrotron radiation itself, in the process known
as synchrotron self-Compton (SSC), or photons from the active nucleus,
particularly at infrared through ultraviolet frequencies.  Since the
available photons range in frequency, so too do the energies of
electrons responsible for scattering them into the X-ray, and these
are rarely the same electrons for which the magnetic field is probed
through synchrotron radiation. Nevertheless, it is usual to assume
that the magnetic field, photons, and relativistic electrons are
co-located, with the synchrotron photon density proportional to
$\kappa B^{1+\alpha}$.  Here $\alpha$ is defined as in
Equation~\ref{eq:freqspectrum}, and theory gives $\alpha = (p-1)/2$.
The combination of synchrotron
(radio) flux density and inverse Compton (X-ray) flux density then
allows a value for the magnetic field strength, $B_{\rm SiC}$, to be
inferred and compared with $B_{\rm me}$.

Since the modelling requires that the volume and any bulk motion of
the emitting plasma be known, the best locations for testing minimum
energy are the radio hotspots, which are relatively bright and
compact, and are thought to arise from sub-relativistic flows at jet
termination \citep[but see][]{georg-dechotspot}, and old radio lobes
where the plasma may be relatively relaxed.  There is no reason to
expect dynamical structures to be at minimum energy.

It was anticipated that \chandra\ and \xmm\ would make important
advances in tests of minimum energy, since already with \rosat\ and
\asca\ there were convincing detections of inverse Compton X-ray
emission from the hotspots and lobes of a handful of sources
\citep[e.g.,][]{harris-cyga, feigelson-loberos, tash-cenb-lobasca}, and
pioneering work on the hotspots of Cygnus~A had found good agreement
with minimum energy \cite{harris-cyga}.  \chandra\ and \xmm\ have
allowed such tests to be made on a significant number of lobes and
hotspots, with results generally finding magnetic field strengths
within a factor of a few of their minimum energy (equipartition)
values for $K = 0$ and $\eta = 1$ \citep[e.g.,][]{hard-3c123hot,
brun-3c207, isobe-3c452lobe, comastri-minen3C219, bondi-minen3C265,
belsole3, croston2, migliori}.  A study of $\sim 40$ hotspot X-ray
detections concludes that the most luminous hotspots tend to be in
good agreement with minimum-energy magnetic fields, whereas in
less-luminous sources the interpretation is complicated by an
additional synchrotron component of X-ray emission
\cite{hard-hotspots2}.  Considerable complexity of structure
is seen where hotspots are close enough for X-ray images to have
kpc-scale or better resolution
\citep[e.g.,][]{kraft-3c33}.  

\begin{figure}[t]
\centering
\includegraphics[height=5.8cm]{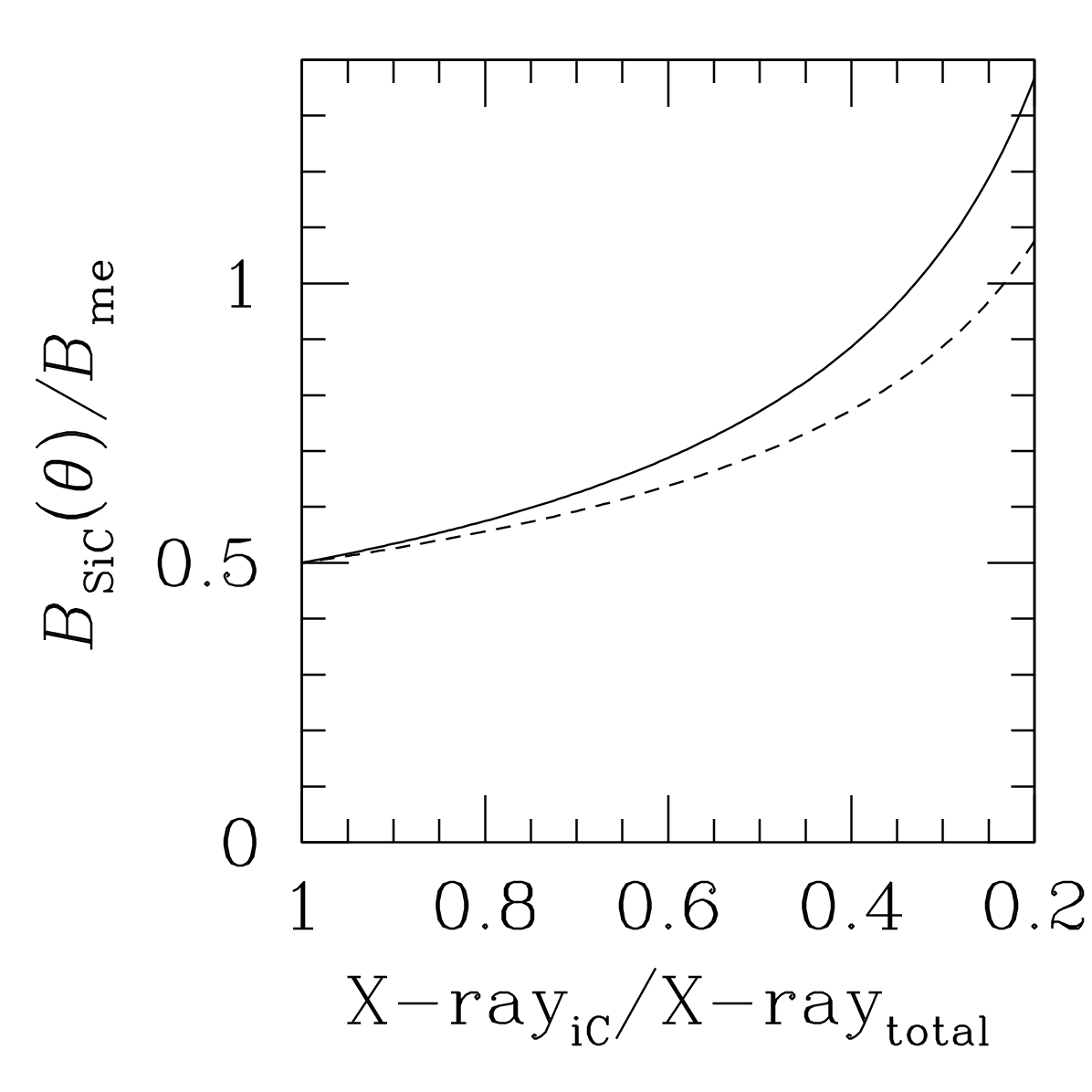}
\caption{The amount by which the fraction of the total X-ray flux
density attributable to inverse Compton radiation, X-ray$_{\rm
iC}$/X-ray$_{\rm total}$, would have to be reduced for a result of
$B_{\rm SiC}/B_{\rm me} = 0.5$ to be increased.  Solid and dashed
curves are for $\alpha=0.6$ and $\alpha=1.1$, respectively.  }
\label{fig:bsic}      
\end{figure}

For radio lobes, 
the largest systematic study where it is assumed
that all the X-ray emission is inverse Compton radiation is of 33 FRII
lobes, and finds $0.3 < B_{\rm SiC}/B_{\rm me} < 1.3 $
\cite{croston-lobes}.  Since the asymmetry is on the side of $B_{\rm
SiC} < B_{\rm me}$, it is important to recognize that the analysis may
not have accurately taken into account contributions to the
lobe X-ray emission from cluster
gas, now commonly detected away from the lobe regions in FRII radio
galaxies \citep[][and see Fig.~\ref{fig:3c200}]{belsole-env}.  However, as seen in
Figure~\ref{fig:bsic}, the lobe X-ray emission from cluster gas would have
to be far brighter than that from inverse Compton scattering to cause $B_{\rm
SiC}/B_{\rm me}$ to increase significantly (e.g., from 0.5 to 1.0), and this
is incompatible with the observation that lobes stand out in X-rays as
compared with adjacent regions.

Better agreement between $B_{\rm SiC}$ and $B_{\rm me}$ would be
achieved if $B_{\rm me}$ has been overestimated.  Figure~\ref{fig:bme}
shows that decreasing the filling factor or including relativistic
protons that energetically dominate the electrons
have the opposite effect.  A
decrease in $B_{\rm me}$ is found if the source has been assumed to be
in the plane of the sky whereas it is really at a small angle, with
the structures having more volume.  However, the small angles required
to make an appreciable difference would be inconsistent with random
sampling.  More promising would be if $\gamma_{\rm min}$ were higher
than typically assumed, as stressed by \cite{blundell} who claim
evidence for a value of $\gamma_{\rm min}$ as high as $\sim 10^4$ in
the hotspot of one FRII radio galaxy, with a lower value of
$\gamma_{\rm min} \sim 10^3$ in the lobes as a result of adiabatic
expansion.  This is in line with earlier measurements of spectral
flattening at low radio frequencies in hotspot spectra,
suggestive of values of $\gamma_{\rm min}$ no lower than a few hundred
\citep[e.g.,][]{leahy89, carilli91}.  Why there
might be such a $\gamma_{\rm min}$ in a hotspot is discussed by
\cite{godfrey}.

\begin{figure}[t]
\centering
\includegraphics[height=5.8cm]{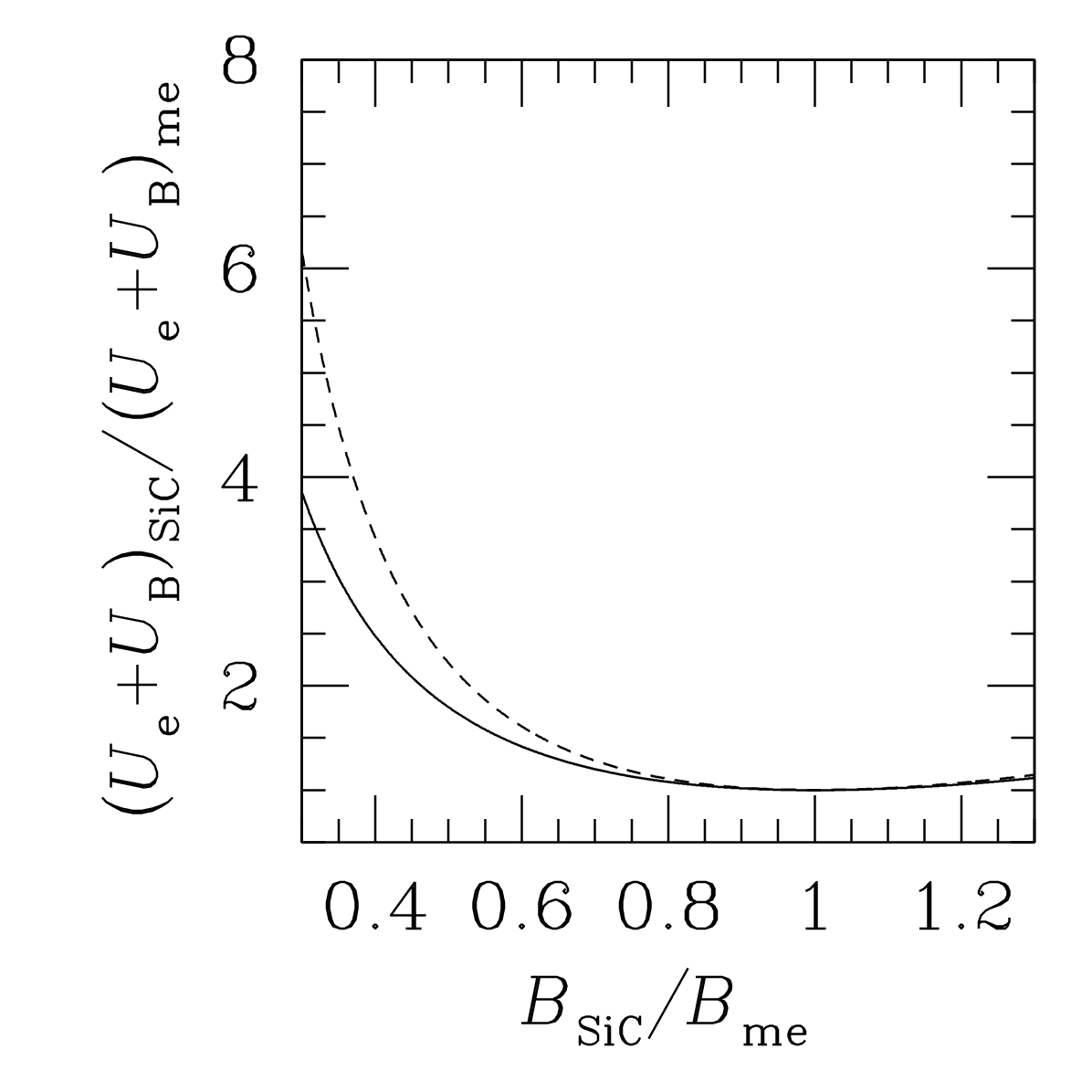}
\caption{The ratio of  
total energy in electrons and magnetic field, computed from
combined X-ray inverse Compton and radio synchrotron measurements, to
that calculated for minimum energy, for the range of $B_{\rm SiC}/B_{\rm me}$
typically observed.  Solid and dashed curves are for
$\alpha=0.6$ and $\alpha=1.1$, respectively.
}
\label{fig:toten}      
\end{figure}

It is important to stress that finding $B_{\rm SiC}/B_{\rm me}$ within
a factor of a few of unity does not allow strong constraints to be
placed on physical parameters.  As shown in Figure~\ref{fig:bme},
large changes in input parameters do not change $B_{\rm me}$, and thus
$B_{\rm SiC}/B_{\rm me}$, by a large amount.  It is often pointed out
that if the magnetic-field strength is a factor of a few below $B_{\rm
me}$, the energy in relativistic electrons must dominate the
magnetic-field energy by orders of magnitude.  While this is relevant
for understanding the state of the plasma, does this really matter
from the point of view of source energetics?  The increase in combined
electron and magnetic-field energy over the minimum energy is
relatively modest as long as the electron spectrum is not very steep
and the field strength is no less than about a third of $B_{\rm me}$
(Fig.~\ref{fig:toten}).

In any case, it is clear that application of
minimum energy over large regions is an oversimplification.
Three-dimensional
magneto-hydrodynamical simulations that incorporate particle transport
and shock acceleration  \cite{tregillis, tregillis2} find
much substructure of particle distributions and fields within the volumes
typically integrated over observationally.  Complexity
on a coarser scale is seen in some observations
\citep[e.g.,][]{isobe-3c452lobe, migliori}.

\section{Do powerful large-scale jets have fast spinal speeds?}
\label{sec:jetspeed}

\subsection{The impetus from PKS 0637-752}
\label{sec:0637}

\begin{figure}[t]
\centering
\includegraphics[height=5.6cm]{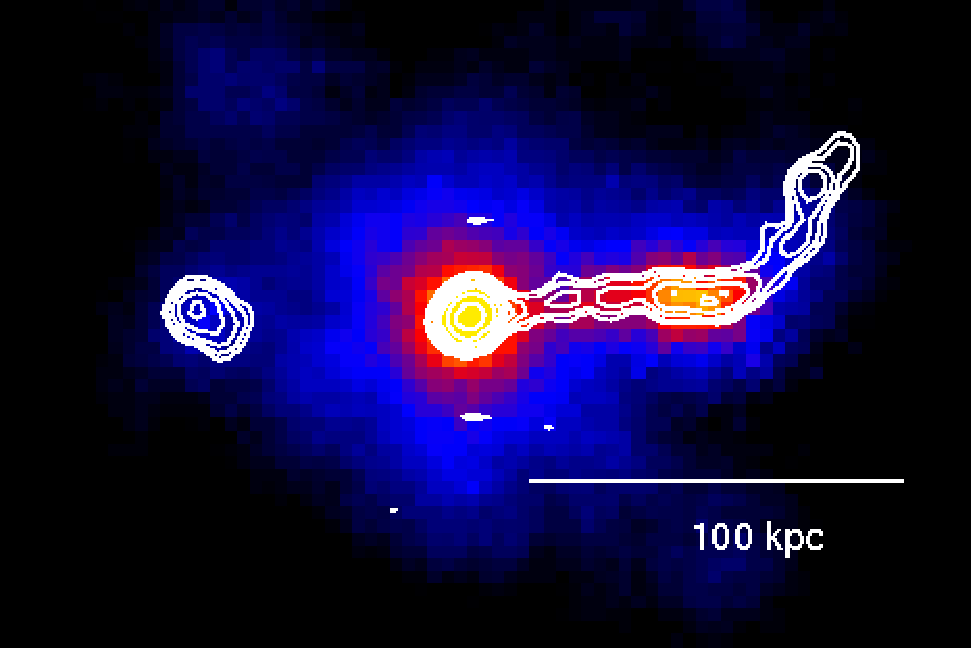}
\caption{The $z=0.651$ quasar PKS~0637-752, using data from
  \protect\cite{schwartz-pks0637}.
The plot shows
a smoothed \chandra\ X-ray image of $\sim 35$~ks exposure with
radio contours
from an 8.64~GHz \atca\ radio map
(beam size $0.96''\times 0.81''$).  X-ray emission is detected from
the nucleus and from the western radio jet before it bends north.
The bright jet region $7.8''$ west of the nucleus is known as
Knot WK7.8.}
\label{fig:pks0637}      
\end{figure}

\chandra\ is central to the current debate concerning jet speed in the
powerful radio jets of quasars.  The work was kick-started
unexpectedly.  Observing quasars was not initialy a high scientific
priority for \chandra, as it was recognized that the cores were
bright, and the likelihood of multiple photons arriving between CCD
readouts was high, leading to distorted spectral measurements (so
called `pileup').  It was thus fortuitous that a radio-loud quasar was
the chosen target for in-flight focus calibration, since this led to
the detection of resolved jet emission from the $z = 0.651$ quasar PKS
0637-752 \citep[][and see Fig.~\ref{fig:pks0637}]{schwartz-pks0637,
chartas-pks0637}.

Several possible origins for PKS 0637-752's jet X-rays were
considered.  The level of optical emission was too low to explain the
X-rays as the synchrotron radiation from a single population of
electrons, and SSC was disfavoured as it would require strong
dominance of the energy in relativistic electrons over that in
magnetic field, giving a total energy in particles and field that is
$\sim 1000$ times that given by minimum energy
\citep{schwartz-pks0637}.  A more promising explanation allowed the
jet to be at minimum energy but required it to have fast bulk motion
(a Lorentz factor of $\Gamma ~\sim 20$ at $\theta \sim 5^\circ$ to the line of sight), in
which case it would see boosted CMB in its rest frame and emit beamed
X-rays in the observer's frame \citep{tav-pks0637, cel-pks0637}.
Although such a speed and angle are consistent with VLBI measurements
on pc scales \citep{lovell-pks0637}, the fast speed must persist up to
hundreds of kpc from the core (after projection is taken into account)
for the X-rays to be produced by this mechanism, which I will call
``beamed iC-CMB''.  This explanation ran counter to
the common wisdom of the time, based on radio data, that the bulk
relativistic speed of quasar jets on the large scale is $\Gamma \sim
2$ \citep[e.g.,][]{bridle, wardle-jetspeed}.  To overcome the contradiction, it
was suggested that quasar jets have a fast-moving central spine
responsible for the observed X-rays, and a slower-moving outer region
that emits the bulk of the observed radio emission  \citep{cel-pks0637}.
This follows the
same pattern as the
transverse velocity structures, 
conjectured for FRI jets, that are thought to result from the
entrainment of external material (see \S\ref{sec:jetcollimate}).

\subsection{The dependence of beamed iC-CMB on beaming factors and redshift}
\label{sec:iC-CMB}

In modelling beamed iC-CMB emission, most authors use the
approximation that CMB photons, isotropic in the observer's frame, are
scattered into directions in the jet frame that are parallel to the
instantaneous velocity vectors of the scattering electrons
\citep[e.g.,][]{dermer, harris-krawic}.  This has been shown to be an
excellent approximation for calculating the X-ray emissivity as long
as the jet's bulk motion has Lorentz factor $\Gamma \geq 2$
\citep{dermersturners}, which is, in any case, required for the
mechanism to be effective at producing strong X-ray fluxes.  The
basic physics of the formalism is particularly clearly presented in
\citep{dermer}, and here those formulae are presented in 
a slightly different form which is independent of the system of units.

We consider a source travelling at
speed  $\beta c$ and bulk Lorentz factor $\Gamma$ 
towards the observer at an angle $\theta$ to the line
of sight, 
so that
the bulk relativistic Doppler factor, $\delta$, is given by

\begin{equation}
\delta = {1\over \Gamma(1-\beta\cos\theta)}.
\label{eq:dopplerfactor}
\end{equation}

An electron of Lorentz factor $\gamma$ will scatter a CMB photon that
has a characteristic frequency today of $\nu_{\rm CMB}$ to an observed frequency,
$\nu$, given by

\begin{equation}
\nu = \nu_{\rm CMB} \gamma^2 {\delta^2 (1 + \cos\theta) \over (1 + \beta)},
\label{eq:gammaGamma}
\end{equation}

\noindent
where the spectral redistribution function is approximated as a delta
function \citep[equation (7) of][written in the notation of this paper]{dermer}.
A delta-function approximation is also
used for the synchrotron spectral distribution function such that
an electron of Lorentz factor $\gamma$ radiates at frequency

\begin{equation}
\nu = \gamma^2 \nu_g ,
\label{eq:gyrofreq}
\end{equation}

\noindent
where $\nu_g$ is the non-relativistic electron gyrofrequency,
which is proportional to the magnetic field strength, $B$.  Written in
SI units, $\nu_{\rm g} = eB/2\pi m_{\rm e} \approx 30 B$~GHz, where
$B$ is in units of Tesla.
For a CMB that is monochromatic at
a frequency of $\nu_{\rm CMB}$ at redshift equal to zero, then 
the ratio of inverse Compton to synchrotron
flux density at a fixed frequency in the observer's frame is simply given by

\begin{equation}
{S_{\rm iC-CMB}\over S_{\rm syn}} = {3\over 4}{\delta^{1+\alpha}}
(1 + z)^{3 + \alpha} \left({1 + \cos\theta \over 1 +
\beta}\right)^{1+\alpha} {u_{\rm CMB} \over u_{\rm B}}
\left(
{\nu_{\rm CMB} \over \nu_g}
\right)^{\alpha - 1},
\label{eq:beamicratio}
\end{equation}

\noindent
where $u_{\rm CMB}$ is the energy-density of the CMB at a redshift of
zero and $ u_{\rm B}$ is the energy density in the magnetic field in
the rest-frame of the jet.
Noting that $u_{\rm B} \propto B_{\rm int}^2$ and
$\nu_g \propto B_{\rm int}$,  where $B_{\rm int}$ is the intrinsic magnetic-field
strength in the rest-frame of the jet,

\begin{equation}
{S_{\rm iC-CMB}\over S_{\rm syn}} \propto {\delta^{1+\alpha}\over
B_{\rm
int}^{1+\alpha}} (1 + z)^{3 + \alpha} \left({1 + \cos\theta \over 1 +
\beta}\right)^{1+\alpha}.
\label{eq:beamic}
\end{equation}

If the modelling assumes minimum energy in relativistic particles and
fields, then Equation~\ref{eq:we6} can be used.  The luminosity density
can be written in terms of the observable synchrotron flux density
using

\begin{equation}
L_\nu   \delta^{(3 + \alpha)}= (1+z)^{\alpha -1}\, S_\nu\, 4 \pi
D_{\rm L}^2,
\label{eq:we7}
\end{equation}

\noindent
where $D_{\rm L}$ is the luminosity distance.  The volume of a
radio
source can be specified in terms of its angular component
sizes, $\theta_{\rm x}$, $\theta_{\rm y}$ and path length through the
source, $d$, as

\begin{equation}
V = \theta_{\rm x}\theta_{\rm y}\, d\, D_{\rm L}^2 / (1 + z)^4.
\label{eq:we8}
\end{equation}

Substituting for $L_\nu$ and $V$ (Equations \ref{eq:we7} and
\ref{eq:we8})
in Equation \ref{eq:we6} then gives

\begin{equation}
B_{\rm me} = \left[{(\alpha + 1) C_1 \over 2 C_2} 
{(1 + K)\over \eta\, \theta_{\rm x}\theta_{\rm y} d} 4\pi\, {S_\nu
\over \delta^{(3 + \alpha)}}\,
\nu^{\alpha}\, (1+z)^{3 + \alpha}
 {\left(
\gamma_{\rm max}^{1 - 2\alpha} - \gamma_{\rm min}^{1 - 2\alpha}
\right)\over ({1 -2\alpha})}\right]^{1/(\alpha + 3)},
\label{eq:we9}
\end{equation}

\noindent
i.e.,

\begin{equation}
B_{\rm me} \propto {(1+z) \over \delta}.
\label{eq:simple}
\end{equation}

Substituting for $B_{\rm int} = B_{\rm me}$ in
Equation \ref{eq:beamic} gives

\begin{equation}
{S_{\rm iC-CMB}\over S_{\rm syn}} \propto \delta^{2+2\alpha} (1 +
z)^{2} \left({1 + \cos\theta \over 1 + \beta}\right)^{1+\alpha}.
\label{eq:beamicminen}
\end{equation}

\noindent
Equation~\ref{eq:we7} (and thus Equations \ref{eq:we9},
\ref{eq:simple} and
\ref{eq:beamicminen}) applies to a spherical blob in which
$S_{\rm syn} \propto \delta^{3+\alpha}$: for a continuous jet where
$S_{\rm syn} \propto \delta^{2+\alpha}$, 
$B_{\rm me} \propto 1/\delta^{(2+\alpha)/(3+\alpha)}$, 
and 
Equation~\ref{eq:beamicminen} has a slightly more complicated dependence on $\delta$.
Also, Equation~\ref{eq:we8} adopts the assumption that the
pathlength through the jet is independent of redshift.
Alternative assumptions could be adopted, modifying the redshift
dependencies in Equations~\ref{eq:we9}, \ref{eq:simple} and ~\ref{eq:beamicminen}.

\subsection{How is the beamed iC-CMB model faring under scrutiny?}
\label{sec:iC-CMBcritic}

It was obvious that there were important consequences
if the beamed iC-CMB interpretation of the X-ray emission from the
resolved jet of PKS~0637-752 is correct, and holds for other
quasar jets.  In particular, increasing $\Gamma$ from the previously
accepted value of $\sim 2$
to $\Gamma \sim 20$ means increasing the
jet power by a factor of $\sim 100$, or more if
cold ions are an important contributor to the jet composition
\citep[see appendix B of][]{schwartz-4jets}.

Programs targeting the resolved radio jets of core-dominated quasars with
\chandra\ followed the work on PKS~0637-752 \citep{samb-qso6,
samb-qso17, marsh-jetsurvey}.  The detection success rate of roughly
50 per cent in relatively short exposures made it clear that
PKS~0637-752 is not an outlier.  Longer \chandra\ observations were
made of some of the X-ray brightest and morphologically most
interesting sources \citep[e.g.,][]{marsh-chandra3c273, samb-3c273,
siem-1127, siem-1508z=4.3, siem-0738, jorstad-0827, jorstad-5jets,
schwartz-4jets, schwartz-1055, schwartz-longjet, tavech-small-large}.
The combination of surveys and long pointed observations have made it
possible to look critically at the application of the beamed iC-CMB
model to these sources.

\begin{figure}[t]
\centering
\includegraphics[height=6.2cm]{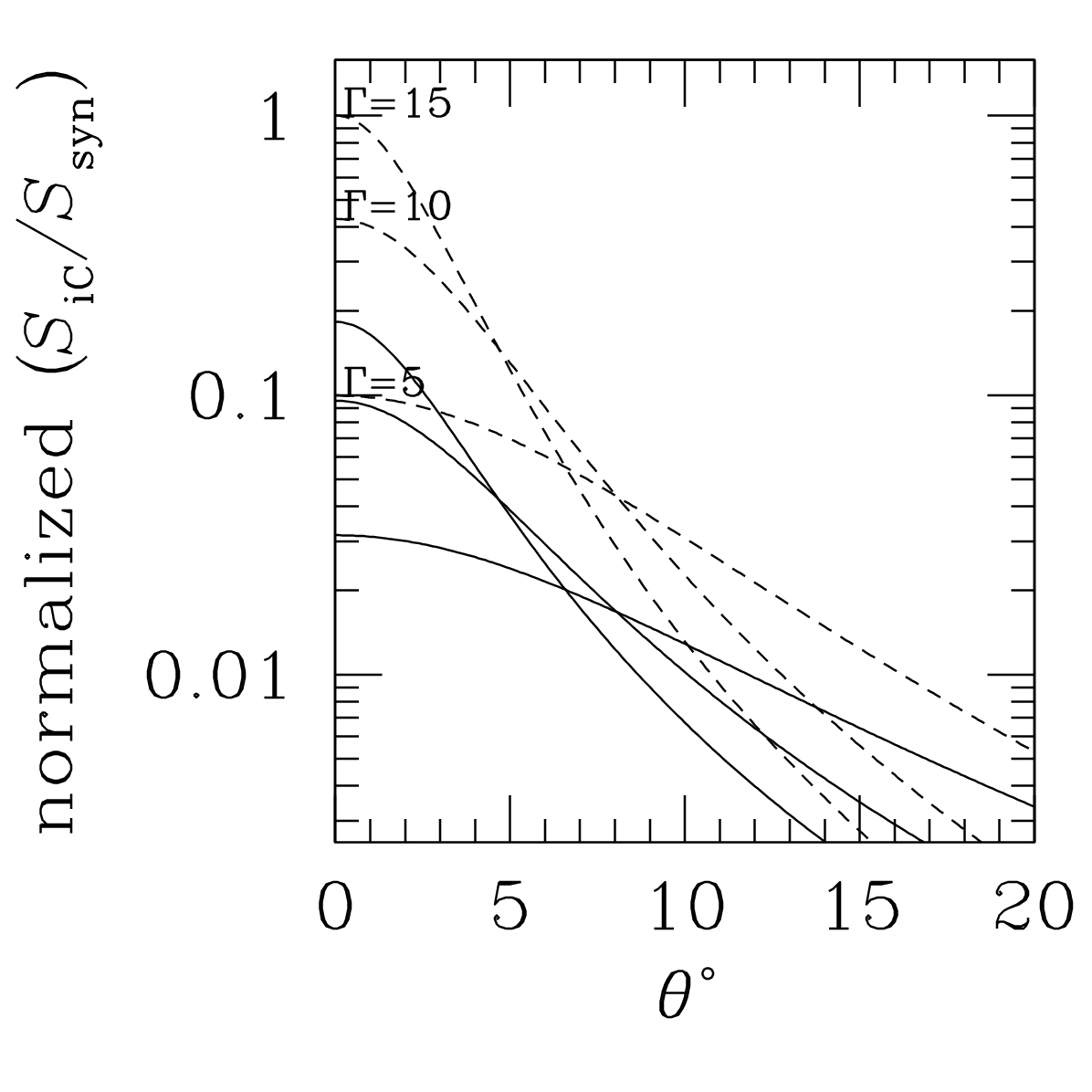}
\caption{Normalized ratio of inverse Compton to synchrotron flux
density for a fixed
intrinsic magnetic field strength for different jet angles to the line
of sight.  The normalization is to the value for a jet with a Lorentz
factor of 15 and $\alpha$=1.1 at $0^\circ$ to the line of sight.
Solid and dashed curves are for
$\alpha=0.6$ and $\alpha=1.1$, respectively.
 Each set has curves for
$\Gamma=15$, $\Gamma=10$, and $\Gamma=5$, in descending order
at $\theta=0^\circ$.
Based on Equation~\ref{eq:beamic}.
}
\label{fig:beamic}      
\end{figure}

The high X-ray detection rate of quasar jets in short exposures is
notable. In most \chandra\ observations of FRII radio galaxies at
similar redshifts to the quasars, the jets (as opposed to the
terminal hotspots) are not detected \citep[e.g.,][]{worr-c220.1,
belsole-env}.  This can be understood in the framework of
quasar/radio-galaxy unification with reference to
Figure~\ref{fig:beamic} (based on Equation~\ref{eq:beamic}) which
shows that for jets that are intrinsically the same, the ratio of
beamed-iC to synchrotron radiation strongly decreases with increasing
jet angle to the line of sight.  The observed quasar X-ray jet
emission is normally one-sided and on the same side as the brighter
radio jet, in support of relativistic beaming.  Where two-sided
X-ray emission has been seen, explanations can be found which are not
in violation of fast jet speeds \citep[e.g.][]{fab-3c9, kataoka}.  

In general the jets contain multiple knots that can be fitted
independently to the beamed iC-CMB model with minimum-energy magnetic
field strengths of order 10--20~$\mu$G (1--2~nT)
\citep[e.g.,][]{schwartz-4jets}.  Note, however, that there are
insufficient observational constraints to fit the two free parameters
of angle to the line of sight and bulk Lorentz factor separately, and
an assumption must be made on one of these parameters.  It has
been common to assume $\sin \theta = 1/\Gamma$ (i.e., $\delta =
\Gamma$), although this is not particularly sensible for sources where
multiple knots in the same source give different values for $\Gamma$,
since it can lead to a jet that bends more erratically than makes
physical sense.  In some cases the results can be shown to agree with
the estimates of speed and power from simple models for the pc-scale
emission \citep[e.g.,][and see \S\ref{sec:beamed}]{jorstad-5jets,
tavech-small-large}, although with rather large uncertainties.

\begin{figure}[t]
\centering
\includegraphics[height=6.4cm]{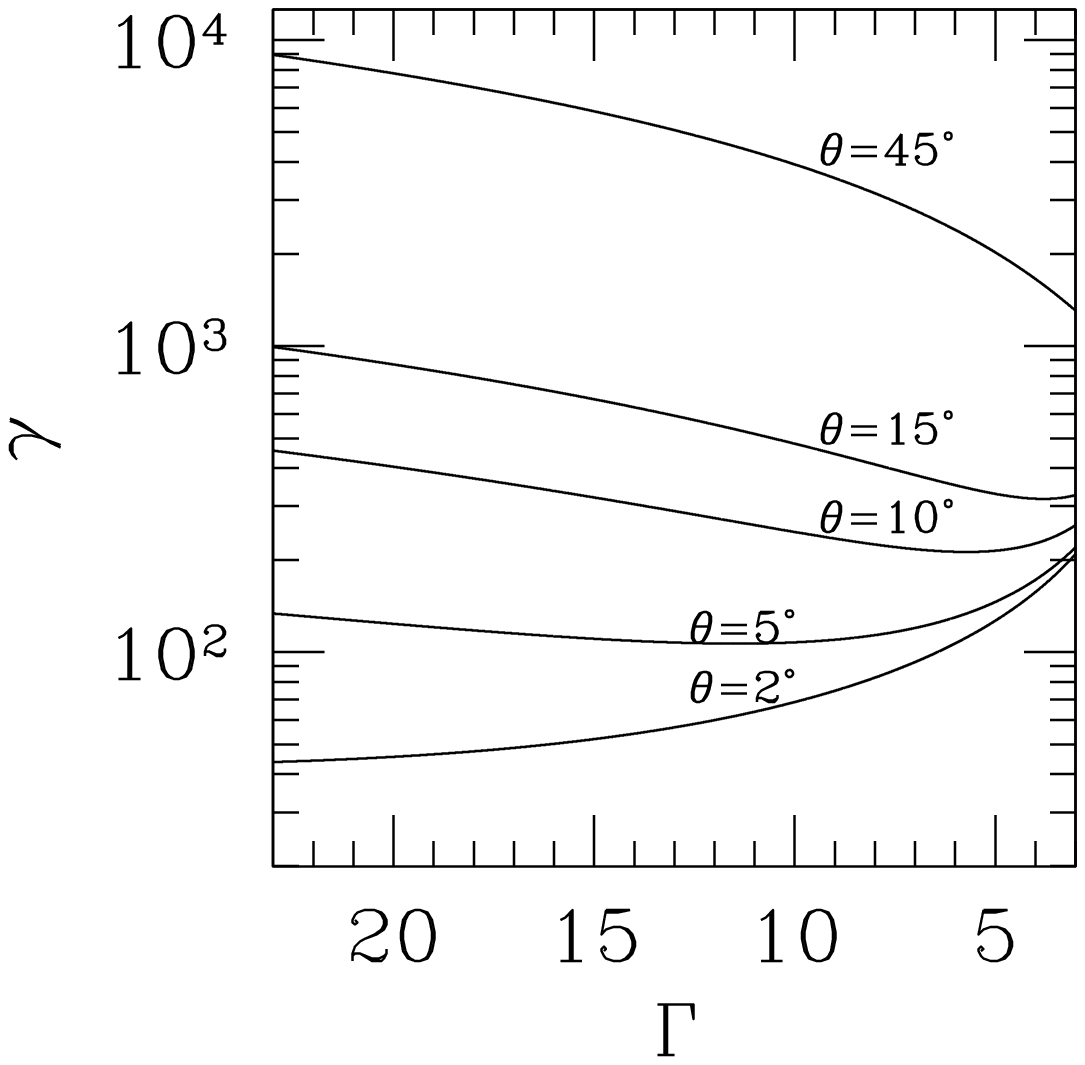}
\caption{Mean Lorentz factor, $\gamma$, of electrons which scatter CMB
photons near the black-body peak to X-ray photons of 1~keV.
Results are shown for an emission region at selected angles to the
line of sight over a range of bulk Lorentz factor,
$\Gamma$.
Based on Equation~\ref{eq:gammaGamma}.
}
\label{fig:gammaGamma}      
\end{figure}

\begin{figure}[t]
\centering
\includegraphics[height=6.8cm]{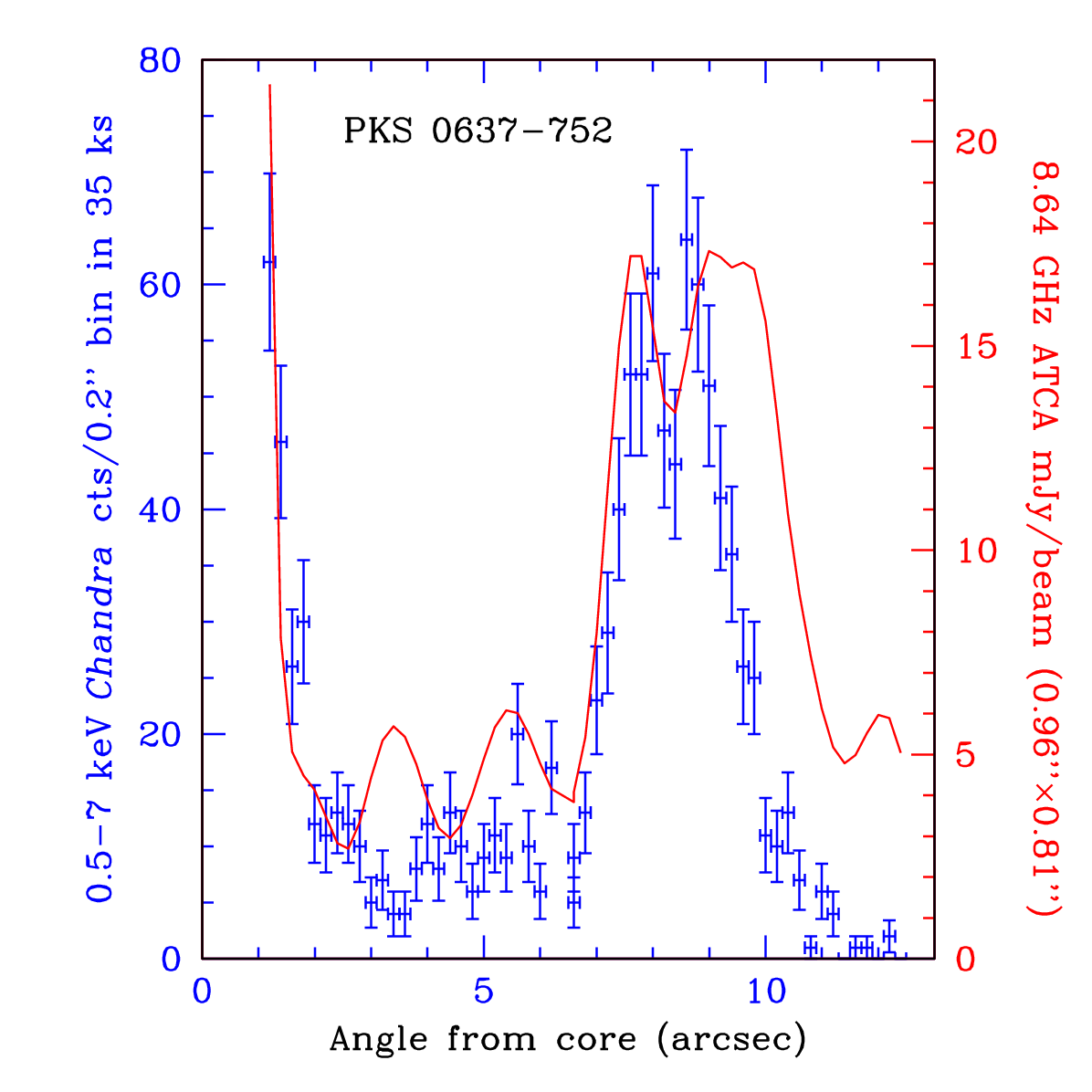}
\caption{The X-ray and radio profiles down the jet of
PKS 0637-752 (see Fig~\ref{fig:pks0637}).
The X-ray intensity drops before the radio at large jet angles.
$1''$ corresponds to a projected linear
  distance of 6.93 kpc.}
\label{fig:pks0637profile}      
\end{figure}

There is, however, a major difficulty with the beamed iC-CMB
interpretation that arises from a detailed comparison between radio
and X-ray emission.  Figure~\ref{fig:gammaGamma} (based on
Equation~\ref{eq:gammaGamma}) shows the mean Lorentz factor of
electrons that scatter photons from the peak of the CMB spectrum into
the X-ray at 1~keV, for various jet bulk Lorentz factors and angles to
the line of sight.  The synchrotron emission from these electrons will
be at a peak frequency of $\approx \gamma^2\nu_{\rm g} \approx
30\gamma^2B~{\rm GHz}$, where $\nu_{\rm g}$ is the gyrofrequency and
$B$ is magnetic field strength in Tesla.  For a typical field of 2~nT,
the radio synchrotron emission from these electrons is at 0.3~MHz if
$\gamma=100$, or 20~MHz if $\gamma=10^3$, both below the observable
radio band.  Under the beamed iC-CMB model, which requires small angle
to the line of sight, $\theta$, to be effective, the X-ray emission thus
arises from lower-energy electrons than the radio emission.  These
electrons have long synchrotron energy-loss lifetimes.  However,
observations sometimes show X-ray emission that weakens relative to
the radio towards the downstream regions of the jets and in some cases
in individual knots, indicating that the population of low-energy
electrons is being depleted more rapidly than the population of
high-energy electrons, contrary to expectations based on radiation losses.
This was seen in PKS 0637-752 \citep[][and see 
Fig.~\ref{fig:pks0637profile}]{schwartz-pks0637,
chartas-pks0637}, and such behaviour is also seen strikingly in several
other sources including 3C~273 \cite{marsh-chandra3c273, samb-3c273},
quasar 0827+243 \cite{jorstad-0827}, PKS~1127-145 \cite{siem-1127} and
PKS~1136-135 \cite{samb-twojets}.  Various suggestions have been made
to overcome the problem within the framework of the beamed iC-CMB
model, but none is uniformly regarded as satisfactory.

\begin{figure}[t]
\centering
\includegraphics[height=5.75cm]{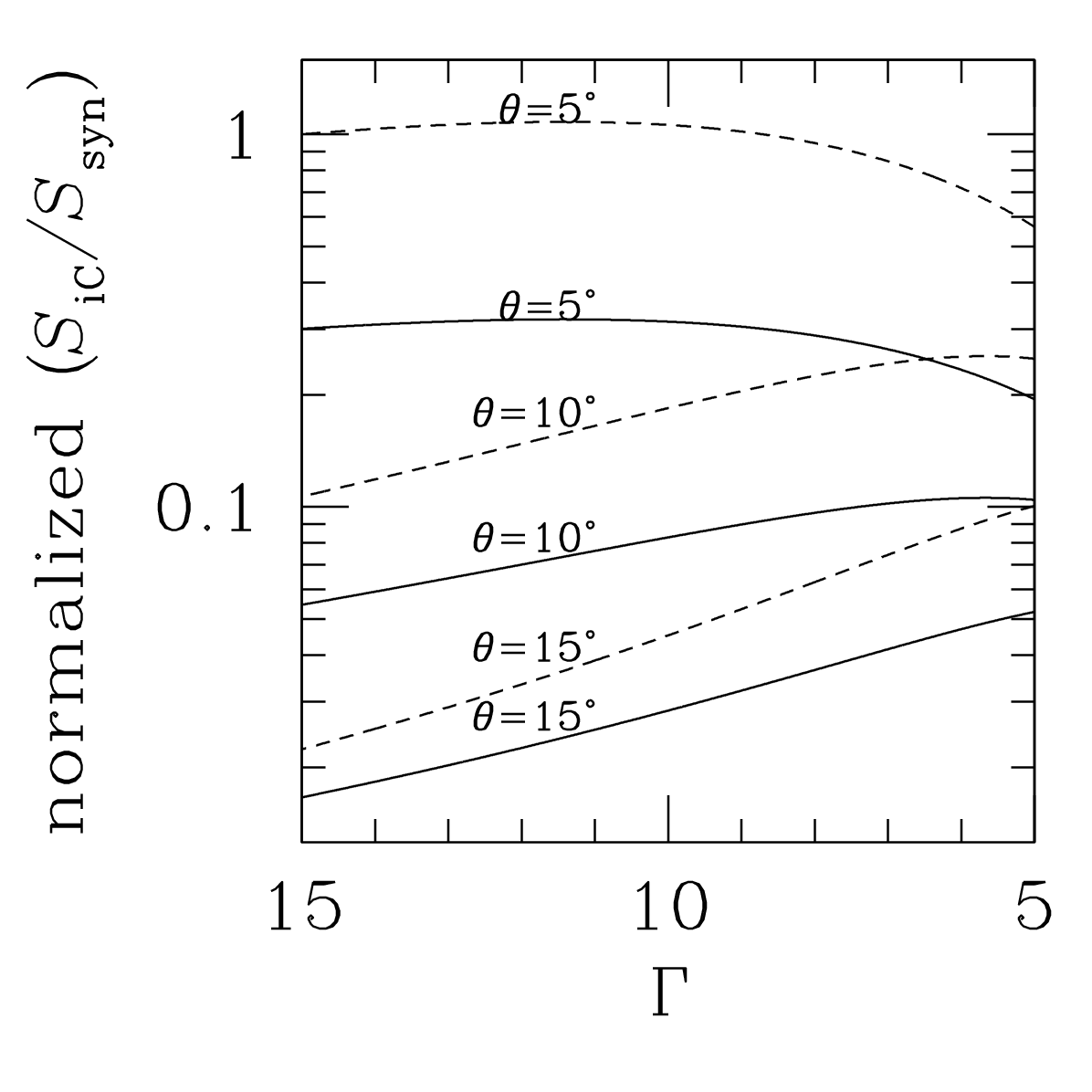}
\includegraphics[height=5.75cm]{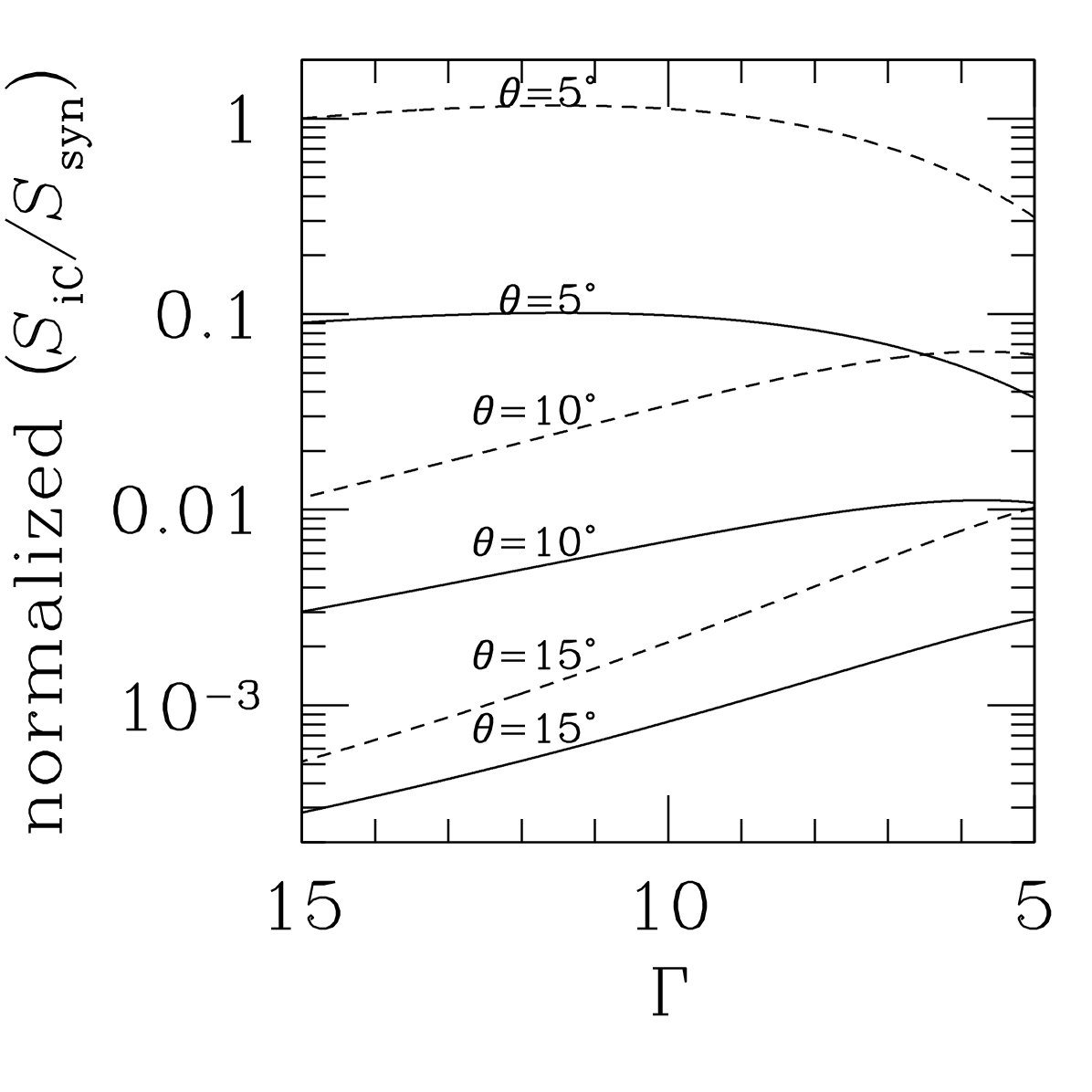}
\caption{Normalized ratio of inverse Compton to synchrotron flux
density for a fixed
jet angle to the line of sight with decreasing
Lorentz factor, $\Gamma$.  Left: with fixed intrinsic magnetic field.
Right: with minimum-energy magnetic field.
In each panel separately the curves are normalized to the
value of $S_{\rm iC}/S_{\rm syn}$ for a jet with a Lorentz
factor of 15, $\alpha$=1.1, at $5^\circ$ to the line of sight.
Solid and dashed curves are for
$\alpha=0.6$ and $\alpha=1.1$, respectively.
 Each set has curves for
$\theta=5^\circ$, $\theta=10^\circ$, and $\theta=15^\circ$, in descending value on the
y-axis at $\Gamma=15$.
}
\label{fig:ics}      
\end{figure}

It has been suggested that strong clumping in the jets may resolve the
problem through adiabatic energy losses \cite{tavech-clumps}.
However, it is not clear that the beamed iC-CMB mechanism is then
required, since such clumping would increase the SSC yield for a slow
jet at minimum energy \cite{schwartz-pks0637}.  Alternatively, it has
been suggested that jet deceleration is important, perhaps through
entrainment of external gas \citep[e.g.,][]{georg-decjet,
samb-twojets, tavech-decel}.  A problem with this as a general
solution is that, as shown in Figure~\ref{fig:ics} (based on
Equations~\ref{eq:beamic} and \ref{eq:beamicminen}), the ratio of
inverse Compton to synchrotron emission only falls for a decelerating
jet over particular ranges of bulk Lorentz factor for jets at an angle
of less than about $5^\circ$ to the line of sight.  This means that
any source for which the X-ray drops off faster than the radio with
downstream distance would need to be at particularly small angle to
the line of sight or rather slow (but see \cite{georg-decjet} for a
more detailed treatment that includes compression of the magnetic
field and thus relative amplification of the radio synchrotron
emission downstream).  Jet deceleration is potentially testable
through looking at the X-ray and radio profiles of source samples.

\begin{figure}[t]
\centering
\includegraphics[height=8.6cm]{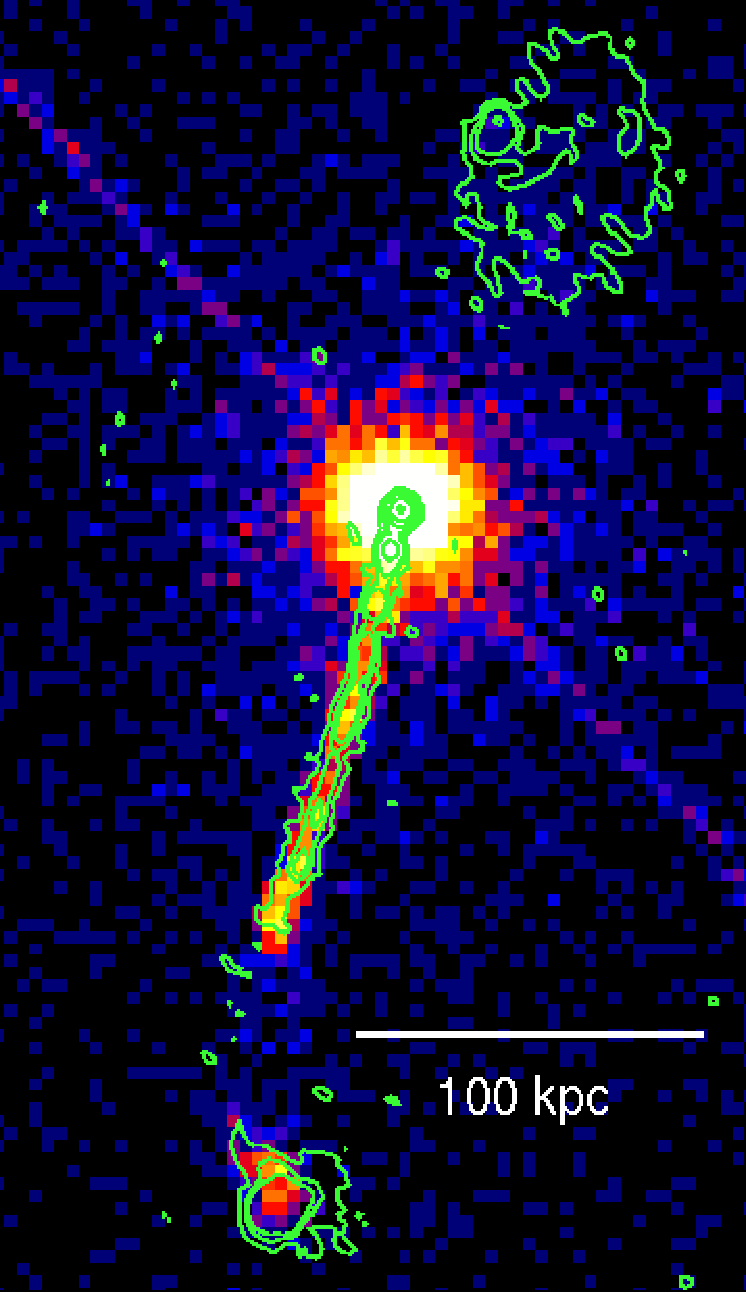}
\caption{The $z=0.72$ quasar 4C19.44, using data from \cite{schwartz-longjet}.
The plot show
an unsmoothed \chandra\ X-ray image of $\sim 189.4$~ks exposure with
radio contours
from a 4.86~GHz \vla\ radio map
(beam size $0.47''\times 0.43''$).  X-ray emission is detected from
the nucleus, the southern radio jet, the northern hotspot and southern
radio lobe. The excess X-ray counts in a line running NE-SW centred on
the nucleus are a frame-readout artifact. }
\label{fig:4c19.44}      
\end{figure}

\begin{figure}[t]
\centering
\includegraphics[height=6.8cm]{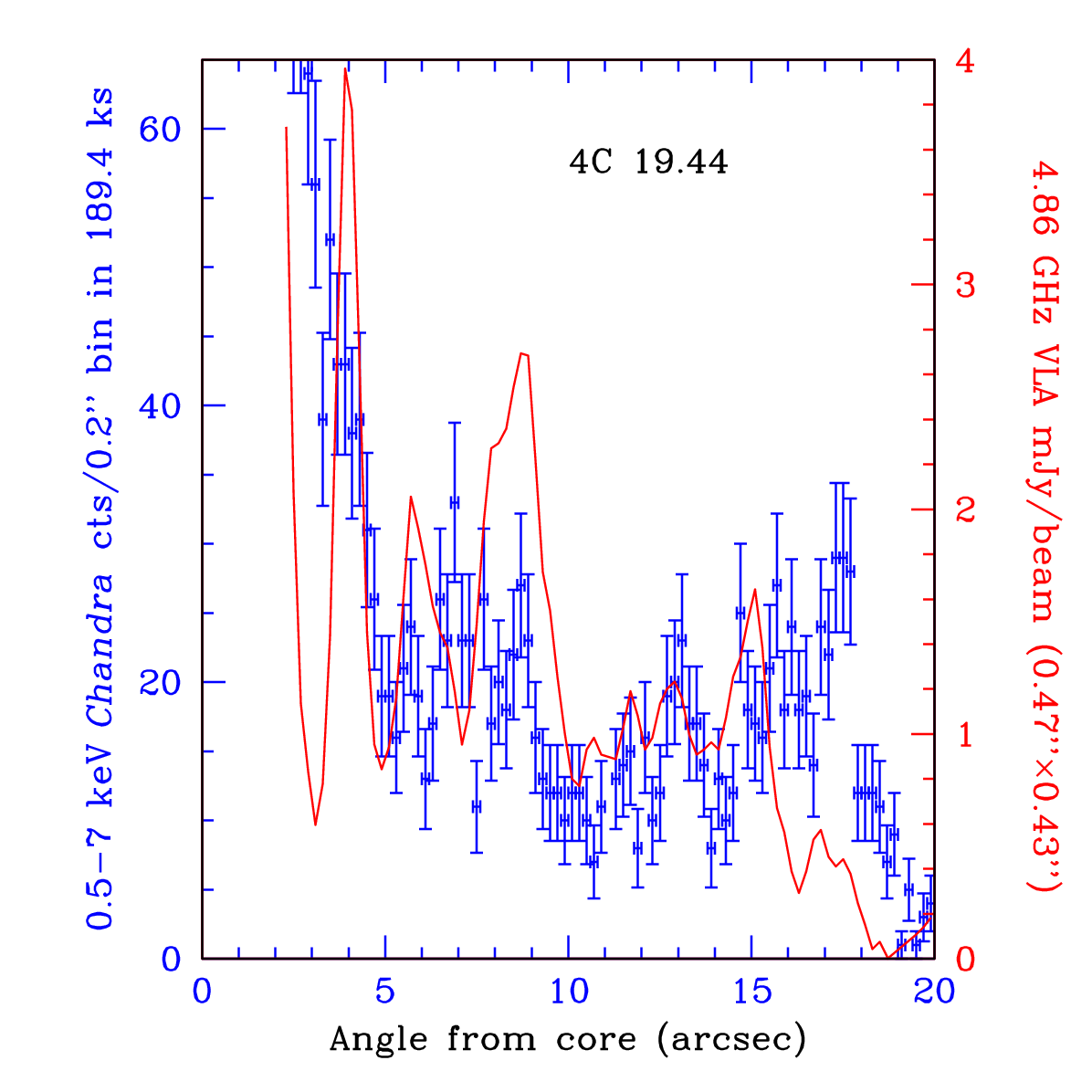}
\caption{The X-ray and radio profiles down the jet of
4C~19.44 (see Fig.~\ref{fig:4c19.44}).  In contrast to PKS~0637-752
(Fig.~\ref{fig:pks0637profile}), the radio intensity drops before the
X-ray at large jet angles.  $1''$ corresponds to a projected linear
distance of 7.23 kpc.}
\label{fig:4c19.44profile}      
\end{figure}

A point in favour of the beamed iC-CMB explanation is that the
particularly straight knotty jet in the quasar 4C~19.44 shows one of
the most uniform X-ray to radio ratios over almost a dozen discrete
knots in its straightest section \citep[][and see
Fig.~\ref{fig:4c19.44}]{schwartz-longjet}.  In contrast to
PKS~0637-752, the radio drops more rapidly than the X-ray at the end
of the straight, well-collimated jet beyond about $15''$ from the
nucleus (Fig.~\ref{fig:4c19.44profile}).  This might suggest that
drops in the level of X-ray to radio emission along other jets are the
result of the jets bending out of the line of sight.  Since quasar
jets are selected for observation based partly on their core radio
emission, any bending downstream is more likely in a direction away
from the line of sight than towards it.  A large change in jet angle
could easily produce the typical decreases in X-ray to radio ratio (a
factor of a few to about 10; compare with Fig.~\ref{fig:beamic}).
However, it is difficult to understand how a real change in angle of a
$\Gamma \sim 20$ flow by more than about a degree could occur without
severe jet decollimation\footnote{Large changes in jet angle in
projection are observed in many sources, but Figure~\ref{fig:beamic}
relates to the true jet angle to the line of sight.}.  As apparent
from Figure \ref{fig:beamic}, more than a factor of about two decrease
in X-ray to radio ratio is then not expected from bending alone.

A test that the beamed iC-CMB explanation must pass concerns the
redshift dependency.  The increase in CMB energy density with redshift
means that the X-ray to radio ratio should increase with redshift by a
factor of something like $(1 +z)^{2}$ (Equation \ref{eq:beamicminen}:
the precise dependence on redshift depends on assumptions concerning
minimum energy and whether or not the path length through the source
is redshift dependent).  Such a redshift effect is not ruled out
\cite{marsh-jetsurvey} although a larger sample is needed for a more
definitive test.

\subsection{Synchrotron emission as an alternative}
\label{sec:iC-CMBsyn}

\begin{figure}[t]
\centering
\includegraphics[height=5.75cm]{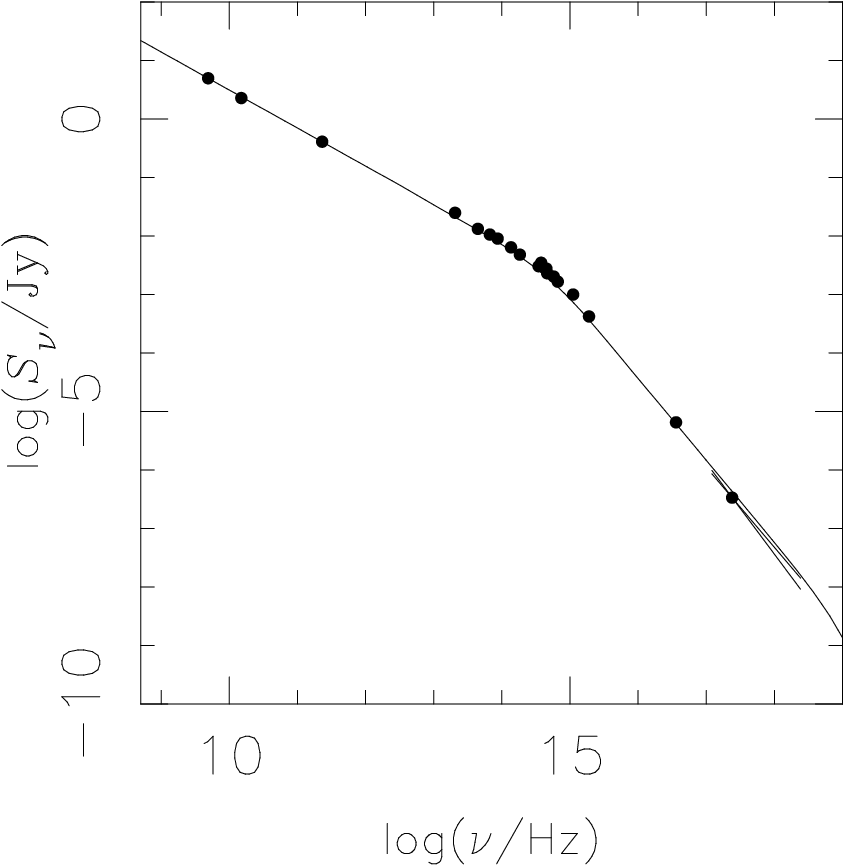}
\includegraphics[height=5.75cm]{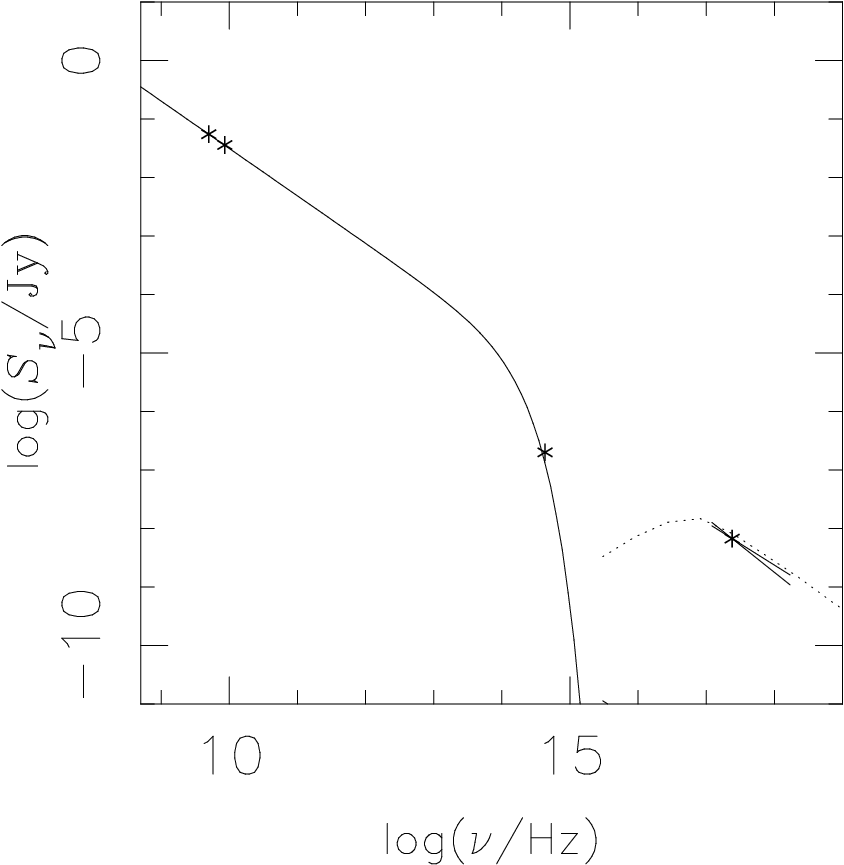}
\caption{Spectral distributions from the radio to X-ray.
Left: The integrated emission from Knots~A, B and C of M~87, using
data from \protect\cite{boer-n1275rosat}
and \protect\cite{berghoefer}, fits a broken power law synchrotron spectrum,
although the change of 1.5 in electron spectral slope is greater than
expected from a simple model for synchrotron energy losses.
Right: A broken power-law spectrum does not
fit through the emission from Knot WK7.8 of PKS~0637-752 (taken from
\protect\cite{chartas-pks0637}) although a synchrotron component with an exponential
cutoff and either a beamed iC-CMB component or a separate synchrotron
component with an anomalously high low-energy cutoff can be made to fit the data.}
\label{fig:sed:m87-0637}      
\end{figure}

The fast jet speed required for the beamed iC-CMB explanation of
quasar X-ray emission disappears if an alternative explanation can be
found for the X-rays.  It is then natural to invoke synchrotron
radiation, the mechanism producing the X-rays in low-power FRI jets
(see \S\ref{sec:Xsyn}).  However, whereas for FRI jets the SED can
normally be modelled with a broken power-law spectrum from the radio,
through the optical to the X-ray \citep[e.g.,][]{boer-m87xmm,
hard-66b, birk-pks0521}, PKS~0637-752 has too little optical emission
to allow this, and a separate population of electrons with an
anomalously high low-energy cutoff would be required
\cite{schwartz-pks0637}.  Figure~\ref{fig:sed:m87-0637} compares the
spectral distribution of the FRI radio galaxy M~87, where a
broken-power-law synchrotron components fits well, with that of the
FRII quasar PKS~0637-752.

Most of the several tens of current quasar X-ray jet detections were
found through targeted \chandra\ programs to observe bright,
prominent, one-sided radio jets.  In most cases there was no
pre-existing reported optical jet detection, but there has been
reasonable success from follow-up work.  The level of such optical
detections often lies below an interpolation between the radio and
X-ray spectra, supporting the idea that synchrotron emission from a
single power-law distribution of electrons is not responsible for all
the emission \citep[e.g.,][]{samb-qso17}.

However, the conclusion regarding synchrotron emission is not quite as
clear cut, since a single-component electron spectrum will harden at
high energies if inverse-Compton losses are also important (since this
loss process is less efficient in the Klein-Nishina regime), and the
consequent spectral hardening in the synchrotron spectrum might then
better match observations \cite{dermer-atoyan}.

\begin{figure}[t]
\centering
\includegraphics[height=5.8cm]{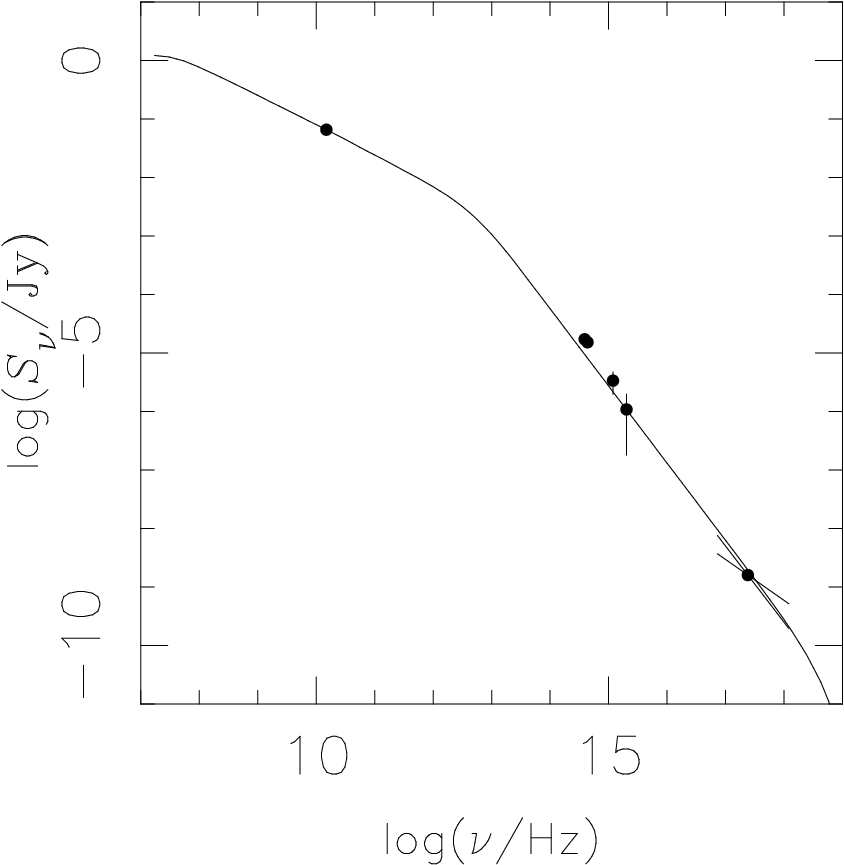}
\caption{The spectrum from the radio to X-ray
for the knot in the FRII radio galaxy 3C~346 (see Fig.~\ref{fig:3c346jet})
fits a broken-power-law synchrotron model.
}
\label{fig:sed-3c346}      
\end{figure}

As long as electrons can be accelerated to high energy (and they can
be in FRIs) they will produce synchrotron radiation at some level.
Radio galaxies are at large angle to the line of sight and any iC-CMB
emission will be beamed out of the line of sight of the observer (an
extension of Fig.~\ref{fig:beamic} to large angle shows that, even for
the most optimistic case, the ratio of iC-CMB to synchrotron emission
drops three orders of magnitude between $\theta=0$ and $\theta =
80^\circ$).  Indeed, synchrotron X-ray emission from knots in the radio
jets of nearby FRII radio galaxies is reported
\citep[e.g.,][]{wilson-pica, worr-3c346, kraft-3c403, kataoka}.  When
optical detections are also available, the energy distributions
\cite{kraft-3c403, worr-3c346} are of similar simple form to those in
FRIs (see Fig.~\ref{fig:sed-3c346}), not requiring the complex
electron spectral forms generally needed to explain quasar X-ray
emission as synchrotron radiation.  Spatial offsets reminiscent of
those seen in FRIs and which are presumably a feature of the particle
acceleration processes (see \S\ref{sec:CenAn315}) are also seen
\cite{worr-3c346, kataoka}.

It remains uncertain as to whether or not in quasars
it is necessary to explain the jet X-ray emission
as the synchrotron output of a distorted electron spectrum 
\cite{dermer-atoyan} or from separate populations of electrons
\citep[e.g.,][]{atoyan-dermer}, as an alternative to the beamed iC-CMB model.
In the case of 3C~273, the run of X-ray spectral slope down the
jet rules out a simple beamed iC-CMB interpretation, but
a two-zone iC-CMB model with a faster spine, although disfavoured,
cannot be ruled out \cite[][]{jester-273}.
If a synchrotron interpretation is sought, similar, simple electron
spectra in all jet regions do not fit observations
\citep[e.g.,][]{roser, samb-3c273, marsh-chandra3c273}.
A two-zone model with faster spine has been proposed,
where, unlike for beamed
iC-CMB in which the X-rays are from the spine, the X-rays would arise from
the shear layer through electron acceleration to very high energy
\cite{jester-273}.  

It is important to understand the primary X-ray emission in quasar
jets, and this remains an observational problem --- more work on
samples and further detailed, deep, multiwavelength observations of
individual sources are needed.  Predictions for yields at higher
energies also differ according to the X-ray emission mechanism, and so
there is a prospect that the new Fermi Gamma-ray Space Telescope will
help in finding solutions \citep[e.g.,][]{dermer-atoyan, georg-3C273}.
Optical polarimetry is potentially a strong discriminant since, unlike
for optical synchrotron emission, the optical emission should be
essentially unpolarized if it is a lower-energy extension of X-ray
emission that is produced via the beamed iC-CMB mechanism
\citep[e.g.,][]{jester-more273, uchiyama}.

\section{What keeps jets collimated?}
\label{sec:jetcollimate}

X-ray measurements of the external medium support arguments that
low-power FRI jets slow through entrainment of this gas.

For the few low-power radio galaxies
with heavily studied, straight, radio jets and counterjets (and so
lying relatively close to the plane of the sky and presumably in relatively
relaxed environments), kinematic models
have been constructed to fit the jet-counterjet asymmetry
\citep{laingbridle-31data, canvin04, canvin05, laingbridle-296data}.
Typically, the jets start fast (relativistic) and relatively faint
with a small opening angle. Then they go through a flaring region where
they steadily broaden and are typically bright both in radio and
X-ray (Fig.~\ref{fig:n315jet}), and finally the opening angle changes
and the jet becomes fainter, particularly at X-ray energies
\citep[e.g.,][]{worr-N315}.  It is in this final region, beyond that
shown for NGC~315 in Figure~\ref{fig:n315jet}, that
the jets are modelled as decelerating steadily 
as they collect mass from the external medium or
stellar winds \cite{komissarov}.  Buoyancy
forces are then important for much of the flow further downstream, 
as the jets adjust to changes in the density of the external medium, causing
deflections from straight-line motion.

In ongoing work, these kinematic models are being extended into
dynamical models, based on conservation laws for mass, momentum, and
energy \cite{bick-n315-cons}, and are being tested for self consistency with the
density and pressure of the external medium.  For one source so far,
3C~31, excellent self consistency has been found \cite{laing-bridle3c31}.
This lends confidence to an understanding of the basic flow behaviour
of these sources.

Deceleration via mass entrainment is consistent with a range of
observational evidence at radio frequencies \cite{laing}, and
naturally leads to the outer parts of the jet (sheath) being
decelerated before the inner (spine).  Applied to more central
regions, the consequence that emission from a slower sheath becomes
relatively more important in jets at larger angle to the line of sight
then resolves difficulties in models that unify BL~Lac objects with
FRI radio galaxies \citep[e.g.,][]{chiaberge00}.

It has been known since the \einstein\ and \rosat\ X-ray observatories
that the minimum pressure in low-power FRI jets (calculated without
relativistic protons) is normally below that of the external
X-ray-emitting medium \citep[e.g.,][]{morg-fripressure,
killeen-n1399conf, feretti-b2rosat, worr-b2atm}.  The model for 3C~31
\cite{laing-bridle3c31} demonstrates that entrainment of the external
medium explains the jet dynamics in the deceleration region, and
pressure balance can be achieved by adding relativistic protons
(with neutrality preserved by balancing proton and electron number
densities) or extending the electron spectrum to lower energies (if
electron-positron charge balance is enforced).  Recent work
\cite{croston-env} has claimed a greater pressure imbalance in FRI
jets that are more in contact with external gas
(less in contact with the plumes or lobes of older jet plasma), and
speculates that the pressure is balanced by heated entrained material,
with an entrainment rate or a heating efficiency that is higher where
jets are in greater direct contact with the X-ray-emitting
atmosphere. This seems in conflict with the entrainment model for
the quasar PKS 1136-135 in the context of the beamed iC-CMB model,
where a standard model would have the jets heavily embedded in old
lobe plasma and yet where the estimated entrainment rate is an order
of magnitude higher than for 3C~31 \cite{tavech-decel}.

While the X-ray-emitting interstellar or intergalactic medium can thus
be controlling the flow where FRI jets are decelerating, and indeed
where buoyancy forces or an excess of gas pressure dominate
\citep[e.g.][]{worr-n326, worr-dead}, FRI radio jets are highly
overpressured in their inner regions close to the nucleus
\cite[e.g.,][]{laing-bridle3c31}.  Here the X-ray emission has yet to
contribute in a significant way to the collimation debate.

The jets of FRII radio galaxies are not significantly in contact with
the external medium for most of their length, so the external medium
is unlikely to control jet collimation, although entrainment of
external gas might be significant over their long propagation paths
\cite{tavech-decel}. Current uncertainties in the jet X-ray emission
mechanism, and thus the particle content and energy, make direct
comparison of the internal and external pressures difficult, except in
the large-scale lobes if dynamical effects are ignored.

\section{Where and how does particle acceleration occur?}
\label{sec:pclaccel}

\subsection{The link with synchrotron X-ray emission}
\label{sec:Xsyn}

\chandra\ found X-ray synchrotron emission to be common in the
resolved kpc-scale jets of FRI radio sources \cite{worr-fr1s}.
The X-ray jets are readily detected in sources covering the whole range
of orientation in unified schemes.  The
several tens of detected sources range from beamed jets 
in BL Lac objects \cite{birk-pks0521, pesce-3c371, samb-s52007} 
to two-sided jets in radio
galaxies \cite{chi-n4261, hard-cena}, with most X-ray jets
corresponding to the brighter radio jet, \citep[e.g.,][]{worr-fr1s, hard-66b,
harris-3c129, marsh-hetgsm87, evans-n6251,
worr-N315}.  Several of the observations have been targeted at sources
already known to have optical jets, from ground-based work or \hst.
However, it's proved easier to detect X-ray jets in modest
{\it Chandra\/} exposures than to detect optical jets in \hst\ snapshot
surveys, because of better contrast with galaxy
emission in the X-ray band than in the optical \cite{worr-fr1s}.

Inverse Compton models for any reasonable photon field suggest an
uncomfortably large departure from a minimum-energy magnetic field in
most low-power X-ray jets \citep[e.g.,][]{hard-66b}, although the
beamed iC-CMB model is a contender for the emission from some BL~Lac
objects \citep[e.g.,][]{samb-s52007}.
Otherwise synchrotron
mission from a single electron population, usually with a broken
power law, is the model of choice to fit the radio, optical, and X-ray
flux densities and the relatively steep X-ray spectra
\citep[e.g.,][]{boer-m87xmm, hard-66b}. 
Given Equation~\ref{eq:gyrofreq}, X-ray synchrotron radiation at 1~keV requires
electrons of energy $\sim 10^{13}$~eV (Lorentz factor $\gamma \approx 2
\times 10^7$)
if the magnetic field strength
is of order 20~nT (200~$\mu$G; the electron energy scales as $B^{-1/2}$).
Averaging over pitch-angle distribution, the lifetime of
synchrotron-emitting electrons is given by

\begin{equation}
\tau = {3 m_{\rm e} c \over 4 \sigma_{\rm T} u_{\rm B} \gamma },
\label{eq:lifetime}
\end{equation}

\noindent
where $m_{\rm e}$ is the electron mass, 
$\sigma_{\rm T}$ is the Thomson cross section, and
$u_{\rm B}$ is the energy density in the magnetic field.
We thus see that electrons emitting 1~keV synchrotron radiation in
a 20~nT magnetic field have an energy-loss lifetime
of about 30 years (lifetime scales as $B^{-3/2}$).
The electrons must therefore
be accelerated {\it in situ}, since their lifetimes against
synchrotron losses are less than the minimum
transport times from the active nuclei, or even from side to side
across the jet.  (This should not be the case if
proton synchrotron radiation is important \cite{aharonian}, since
lifetime scales as $(m_{\rm p}/m_{\rm e})^{5/2}$.)  Particle
acceleration is generally discussed for the cases of a particle
interacting with a distributed population of plasma waves or
magnetohydrodynamic turbulence, or shock acceleration \citep[see
e.g.,][]{blandford-eichler, eilek, hoshino, amato-arons}.

For electrons, particle acceleration and energy losses are in
competition \citep[e.g.,][]{heavens}, no more so than in hotspots of
FRIIs \citep[e.g.,][]{brun-hotspots}, which mark the termination
points of the beam.  Hotspots display considerable complexity in the
X-ray, with synchrotron components seen in the less powerful sources
indicating that TeV electrons are present
\citep[e.g.,][]{hard-hotspots2, kraft-3c33}.  It has been suggested
that the low-energy radio spectral-slope change seen in hotspots may
mark a transition between electrons that are accelerated through 
electron-proton cyclotron resonance and
those (at higher energy) that are simply undergoing shock acceleration
\citep[e.g.,][]{stawarz-cygnusa, godfrey}.  If in FRIs the far-IR
spectral break consistently maps electrons of a particular energy, it
is possible that the break here is also more related to acceleration
than loss processes \cite{birk-spitzer}.

Whether or not particle
acceleration is required along the jets of quasars depends on the
emission process at high energies.  If the beamed iC-CMB model holds,
then the electrons participating in radiation at wavelengths currently
mapped are generally of low enough energy to reach the end of the jet
without significant energy loss, except if a relatively high level of
optical emission must be explained as synchrotron radiation.  The knotty nature
could then be understood as variable output in the jet
\citep[e.g.,][]{stawarz}.  However, in nearby
FRII radio-galaxy jets, where synchrotron X-ray emission is seen
(\S\ref{sec:iC-CMBsyn}), the need for particle acceleration is secure,
and similar underlying processes are expected in quasars even where
the synchrotron X-rays might be outshone by beamed iC-CMB emission.

\begin{figure}[t]
\centering
\includegraphics[height=2.8cm]{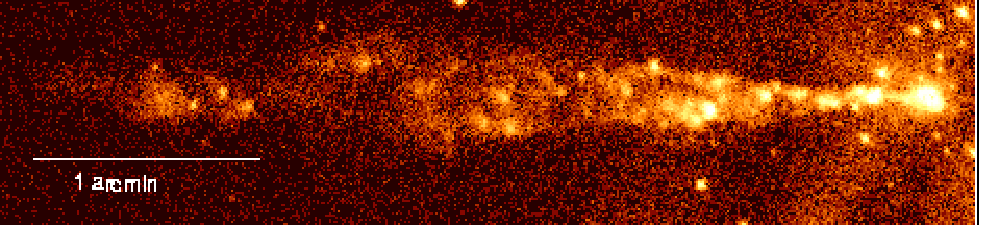}
\caption{A rotated image of a roughly 4.5 kpc (projected) length 
of the 0.8--3~keV X-ray jet of
Cen~A from combining six deep ($\sim 100$~ks) {\it Chandra\/}
exposures.  Image taken from \protect\cite{worr-cenarecent}.}
\label{fig:cenajet}      
\end{figure}

Details of the regions of particle acceleration are best studied in
the closest sources.  Cen~A (Fig.~\ref{fig:cenajet}) and NGC~315
(Fig.~\ref{fig:n315jet}) are
particularly good examples of FRI jets where the X-ray jet emission is
resolved across as well as along the jet, and X-ray
knots are embedded in more diffuse structure \cite{hard-cena, worr-N315,
hard-cenarecent, worr-cenarecent}.  The fact that the X-ray emission
is not just confined to regions within energy-loss light travel
distances of the knots shows that particle acceleration can occur
also in diffuse regions.  The relatively soft X-ray spectrum seen in
the diffuse emission in Cen~A has been used to argue that something
other than shock acceleration (proposed for the knots)
might be taking place in the diffuse
regions \cite{hard-cenarecent}, although no specific explanation is
suggested, and the competition between energy losses and acceleration
may be more important here.

\subsection{Particle acceleration in knotty structures}
\label{sec:CenAn315}

The model of jet deceleration through entrainment
(\S\ref{sec:jetcollimate}) leaves unanswered important questions about
the origins of the bright knots that appear in many jets, particularly
FRIs, and that are usually interpreted as the sites of strong shocks.  Radio
studies have searched for high-speed knot motions, with apparent
speeds greater than the speed of light having been noted in M~87
\cite{biretta}.  A proper-motion study of the knots in Cen~A over a
10-year baseline found that some knots, and even some more diffuse
emission, travel at about 0.5c, indicative of bulk motion rather than pattern
speed \cite{hard-cena}.  This motion, coupled with the jet-to-counter
jet asymmetry, suggests considerable intrinsic differences in the two
jets, to avoid the jets being at an implausibly small angle to the
line of sight.  

Other knots in Cen~A appear to be stationary, which might suggest
that they result from intruders in the flow, such as gas clouds or
high-mass stars \citep[e.g.,][]{fedorenko, hard-cena}.
Some of these  have emission
profiles in the X-ray and radio that are unexpected from a simple toy
model where the electrons are accelerated and then advect down the
jet, losing energy from synchrotron radiation.  Instead the bulk of
the radio emission peaks downstream from the X-ray within these knots,
leading to suggestions that both radio and X-ray-emitting electrons
are accelerated in the standing shock of a stationary obstacle, and a
wake downstream causes further acceleration of the low-energy,
radio-emitting, electrons \cite{hard-cena}.
The resulting radio-X-ray offsets,
averaged over several knots, could give the radio-X-ray offsets
commonly seen in more distant jets \citep[e.g.,][]{hard-66b,
worr-3c346, dulwich-3c15}.  

The knots of Cen~A are not highly variable in observations to date
\cite{hard-cenarecent}, but dramatic variability on a timescale of
months is seen in a knot in the jet of M~87, and the X-ray, optical
and radio light curves are broadly consistent with shock acceleration,
expansion, and energy losses, although the timeline is currently too
short for strong conclusions to be drawn \cite{harris-hstvarM87}.

It is important to study the location of jet knots within the flow, to
see if that can provide a clue as to their nature.  A particularly
interesting example is NGC~315 \cite{worr-N315}.  Here the diffuse
emission contains a knotty structure in the radio and X-ray that
appears to describe an oscillatory filament (Fig.~\ref{fig:n315jet}).  
Although the structure
could be the result of a chance superposition of non-axisymmetric
knots, the level of coherence led to suggestions that the knots might
be predominantly a surface feature residing in the shear layer between
the fast spine and slower, outer, sheath plasma.  If this
interpretation is correct, we might expect the X-ray spectra of the
knots to be similar across the transverse width of the jet.  However,
the distinct knotty emission is only about 10\% of the total in X-rays
and radio along the $\sim 2.5$ kpc of projected jet length over which
it is detected, and with a source distance of $\sim 70$~Mpc the
observations did not allow the spectra of the knot and diffuse
emission to be separated.  

At 3.7~Mpc, Centaurus~A is a much closer example of an FRI radio
galaxy whose knots and diffuse emission are seen over a similar
projected linear distance to that of NGC~315.  An X-ray spectral study of Cen~A's knots
found a spectral steepening with increasing lateral distance from the
jet axis, disfavouring these knots all residing in a shear layer
\cite{worr-cenarecent}.  A flatter X-ray spectrum is seen more central
to the flow, and an alternative explanation to acceleration in
stationary shocks is that the knots here might be formed by stronger
turbulent cascades with more efficient particle acceleration. Knot
migration under the influence of the shear flow might then be
expected, and proper-motion studies might then distinguish between
this interpretation and stationary shocks from stellar or gaseous
intruders entering the flow \cite{worr-cenarecent}.

\subsection{Incorporating polarization data}
\label{sec:FRIpolariz}

There are no current X-ray missions with polarization capabilities.
However, the radio and optical bands probe electron populations
responsible for the X-ray emission, albeit at different electron energies.
If the emission is synchrotron, polarization data provide our best
handle on the direction and relative degree of alignment of
the magnetic field.  Radio observations show that the
fields are relatively well ordered, although there is much complexity.
Broadly, the magnetic fields in FRII jets tend to be parallel to
the jet axis, whereas in FRI jets
they are either predominantly perpendicular, or
perpendicular at the jet centre and parallel near the edges, with the
mixed configurations pointing to perpendicular fields associated
with shocks and parallel fields from shear or oblique shocks \cite{bridle-perley}. 

Optical polarization measurements of resolved jet structures have been
made with \hst.  So far these have mostly concentrated on nearby FRI
radio galaxies, where the optical features are brighter and the
emission mechanism is synchrotron radiation \citep[for an atlas of
polarization images see][]{perl-atlas}.  Work is under way to explore
optical polarization in the jets of FRII radio galaxies and quasars.
As mentioned in \S\ref{sec:iC-CMBsyn}, the optical emission should be
essentially unpolarized if it is an extension of a beamed iC-CMB X-ray
component, in contrast to being of synchrotron origin.

The first jet to be studied in detail in both its radio and optical
polarized emission was M~87,
where there is evidence for strong shock acceleration in compressed transverse magnetic
fields at the base of bright emitting regions, although the
polarization fraction becomes low at the flux maxima
\cite{m87-hstpolarim}. 
Significant differences between the polarization structures seen
in the optical and radio suggest that the sites of acceleration
are different for different electron energies, with the strongest
shocks, that provide acceleration to the highest energies, appearing
in the most central parts of the jet \cite{m87-hstpolarim}.
Detailed work on 3C~15 shows
a jet that narrows from the radio to the optical to the X-ray,
showing that acceleration to the highest energies occurs more
centrally to the flow,
and a mixture of strong shocks and stratified flows can account for the
broad features seen in the optical and radio polarization
\cite{dulwich-3c15}.

\begin{figure}[t]
\centering
\includegraphics[width=5.2cm]{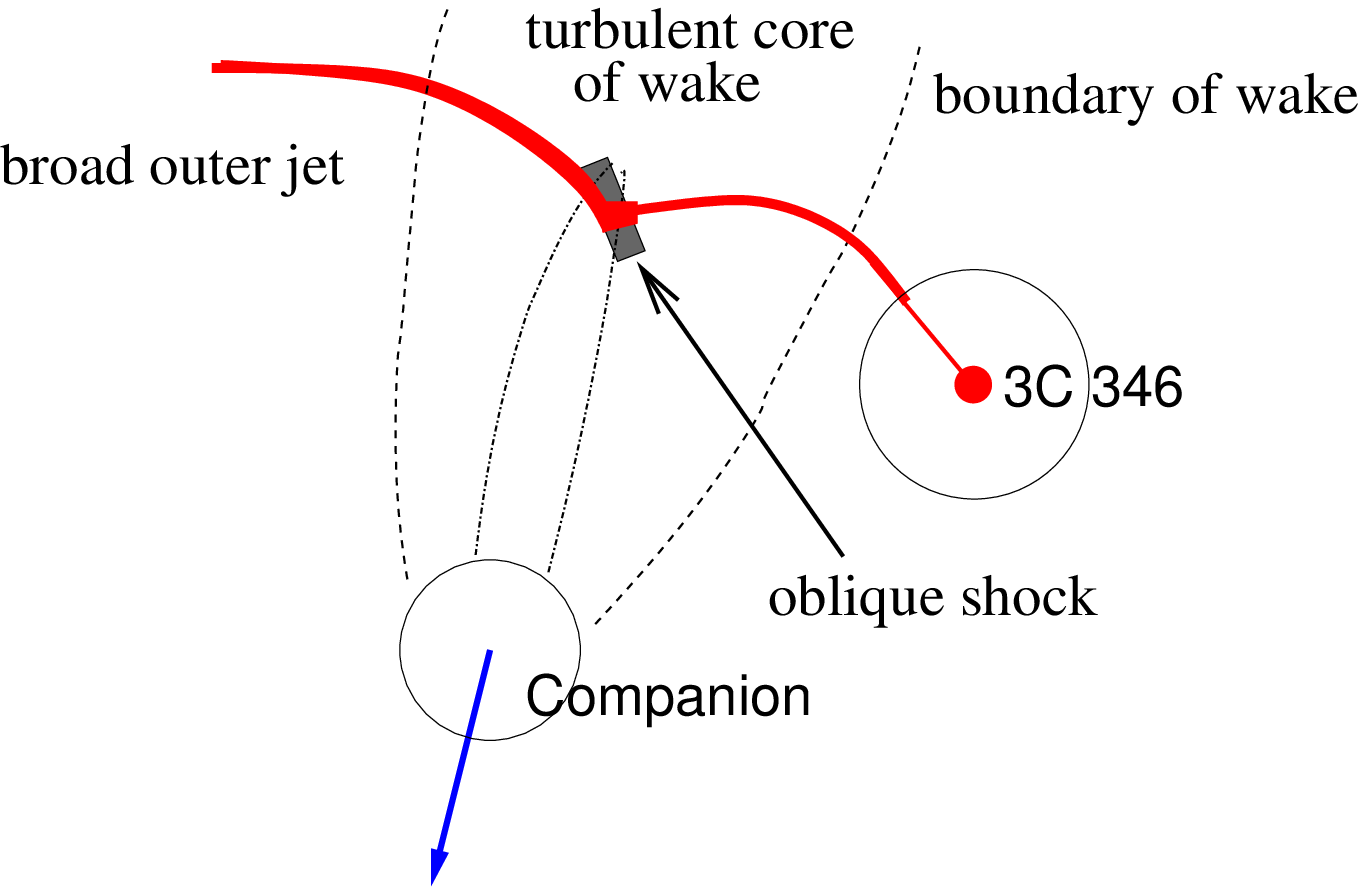}
\includegraphics[height=5.6cm]{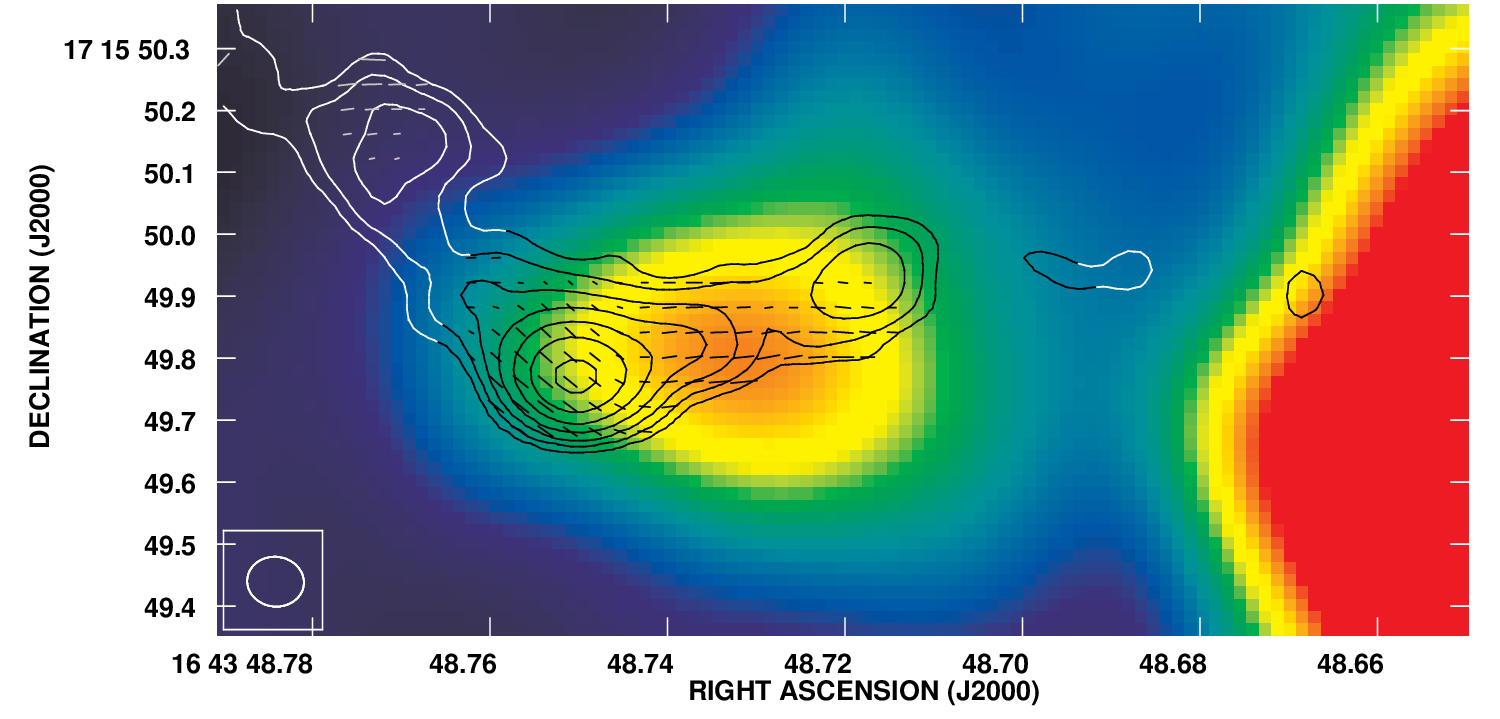}
\caption{3C~346.
Upper: Schematic showing an oblique shock formed in the wake of the
passage of a companion galaxy to 3C~346, and how it affects the radio jet,
from \protect\cite{worr-3c346}.
Circles are the galaxies and red marks the path of the radio jet.
Lower: Radio intensity contours and polarization vectors (rotated
through $90^{\circ}$ roughly to represent the magnetic-field
direction)
on a smoothed \chandra\ X-ray image, indicating compressed
field lines aligned with the proposed shock, from \protect\cite{dulwich-3c346}.
}
\label{fig:3c346jet}      
\end{figure}

A third source for which optical and radio polarization data have been
important is 3C~346 (Fig.~\ref{fig:3c346jet}).  Here X-ray emission is
associated with a bright radio and optical knot where the jet bends by
$70^\circ$ in projection (the X-ray emission peaks somewhat upstream
of the radio, as seen in other sources), leading to a suggestion that
the bending and X-ray brightening are the result of a strong oblique
shock located in the wake of a companion galaxy \cite{worr-3c346}.
Polarization data has supported the model by revealing a compressed
and amplified magnetic field in a direction consistent with that of
the proposed shock, in both the radio and optical \citep[][and see
Fig.~\ref{fig:3c346jet}]{dulwich-3c346}.

 \section{What is jet plasma made of?}
\label{sec:jetcomposition}

Jets are presumed to obtain much of their energy from the infall of
matter into a supermassive black hole.  It is then natural to suppose
that electromagnetic radiation would carry much of the energy from the
system on the smallest scales, since a plausible mechanism for the
extraction of energy is the twisting of magnetic field linked to the
accretion disk \citep[e.g.,][]{lovelace}.  Fast interactions with the
plasma environment and efficient particle acceleration should
load the field with matter.  In the resulting
magnetohydrodynamic flow much of the momentum would be carried by
particles, although Poynting flux may carry a significant fraction of
the total energy \cite{rees71, appl}.

Polarized radiation is the signpost to significant energy in
relativistic particles and magnetic fields.  Jet plasma must be
neutral, on average, to remain collimated, but this can be achieved by various
combinations of relativistic and cold electrons, positrons, and
protons.  Alternatively, it has been suggested that some of the energy is
transported in a decaying neutral beam of ultra-high-energy neutrons
and $\gamma$-rays \cite{atoyan-dermer}.

Several quantities are available to help sort out the
jet composition.

\begin{enumerate}

\item The synchrotron emission.
Since the electron rest mass is only 1/1836 that of 
a proton, and since synchrotron energy loss rates are proportional to
the inverse square of mass,  the observation of synchrotron radiation
is usually used to infer the presence of relativistic electrons
(and perhaps positrons), although
an alternative model produces the synchrotron radiation from
protons accelerated to energies greater than $\sim 10^{18}$~eV
\cite{aharonian}.  

\item The jet power.  All the particles, relativistic and thermal,
combine with the magnetic field strength and bulk Lorentz factor to
produce this quantity \citep[see appendix B of][]{schwartz-4jets}.
It should be no smaller than the radiative power of the old lobe
material (the energy sink), averaged over the lifetime of the source.
In cases where jets have excavated cavities in the external gas, the
enthalpy can be estimated as that required to displace the gas
\citep[e.g.,][]{birzan, dunn, allen}.

\item Faraday rotation.
The contribution from thermal particles must not be so high as
to exceed constraints placed by Faraday rotation, or by
Faraday depolarization for extended regions.  

\item The jet pressure.  Relativistic particles and magnetic field
are thought to dominate this quantity, which is 1/3 of their energy
density, and which can be compared to the external gas pressure.  If
X-ray inverse Compton emission is observed the internal energy density
can be estimated using the radio synchrotron and X-ray flux densities
(\S\ref{sec:minenergytest}).  Otherwise it is usual to assume minimum
energy (\S\ref{sec:minenergycalc}).  A difficulty is that relativistic
electron-proton and electron-positron jets give similar pressures
with different assumptions about the least energetic
particles, for which observational constraints are poor at best.
The contribution of thermal particles to the pressure is usually
taken to be small.

\end{enumerate}

\noindent
Radiation drag and
observational constraints on Comptonized radiation by cold electrons
and positrons seriously hamper electron-positron jets formed close to
the central black hole \cite{sikora, sikora-compton}.  In the cores of
some quasars the radiated power is too large to be met by that
contained in a jet of magnetic field and relativistic leptons close to
minimum energy, and observational constraints on Comptonized radiation
limit the density of cold leptons, so that a significant proton
component is required if the energy carrier is indeed particles
\cite{tav-qsocoresax}.  Thus, an electron-proton plasma is usually
favoured when jets are discussed.  

The presence of relativistic
protons is supported for some FRI radio galaxies: the lobes, if
assumed to be lepton-dominated and radiating at minimum energy
(\S\ref{sec:minenergytest}) would collapse under the pressure of the
X-ray-emitting medium unless there is an additional pressure source
and, although there are several ways of boosting the internal
pressure in such a situation, magnetic dominance would make the
sources unusual, electron dominance is unlikely from constraints on
inverse-Compton scattering of the CMB, and non-relativistic protons
are disfavoured on grounds of Faraday rotation, leaving a relativistic
proton component most likely \citep[e.g.,][]{crost-66b}.  However,
decreased filling factors cannot be ruled out \citep[e.g.,][]{dunn},
except perhaps where the radio structure has excavated a clear
cavity in the X-ray-emitting atmosphere \citep[e.g.,][]{birzan2}.
If indeed the extra pressure is from relativistic protons, it is
uncertain as to how much arises from entrained material accelerated in
the shear layer of the decelerating jet as compared to particles
transported from the core (see \S\ref{sec:jetcollimate}).

FRII jets transport more energy to larger distances, and thus have
more need than FRI jets for a non-radiating energy carrier with high momentum
transport.  Relativistic hydrodynamic simulations find that the key
parameter in preventing jets from strongly decelerating in an external
boundary layer is density contrast with the external medium, in the
sense that denser jets can propagate further \cite{rossi}.  A more
dominant relativistic proton content could provide this.  
Protons are also required if the low-frequency spectral turn-over in
hotspots is the result of cyclotron resonant absorption
\citep[e.g.,][]{godfrey}.
On the other
hand, pressure balance has been used to argue against relativistic
protons in some FRII lobes.  It is argued that the presence of
relativistic protons is improbable since (a) the lobe magnetic field
based on synchrotron and inverse Compton emission agrees with that
from minimum energy calculated using relativistic leptons alone, and
(b) that, even in the absence of such protons, the source is
in pressure balance with the external
medium \citep[e.g.,][]{belsole3, croston2}.  However, these
calculations ignore possible dynamical effects in FRII lobes, and
there are considerable additional sources of uncertainty (see
\S\ref{sec:minenergytest} and item 4 above).

In the context of the beamed iC-CMB model for quasar jets
(\S\ref{sec:iC-CMB}), it is possible to extend an argument limiting
the density of cold electron-positron pairs \cite{sikora-compton} to
kpc-scale regions \cite{georg}.  In the case of PKS 0637-752, upper
limits on Comptonized CMB radiation from \spitzer\ are sufficiently
low to place stringent limits on the mass flux carried by cold lepton
pairs, with the implication that this jet is indeed made electrically
neutral through a strong presence of protons \cite{uchiyama}. However,
this argument relies on the beamed iC-CMB model being correct, with a
large kinematic power being sustained throughout the jet (see
\S\ref{sec:iC-CMBcritic}).

Jet composition remains uncertain, and various degeneracies between
physical quantities and observable parameters render it difficult to
make watertight arguments.  However, X-ray measurements continue to
provide important clues to the puzzle.

\section{What
does X-ray emission tell us about the dynamics and energetics of radio
plasma/gas interactions?}
\label{sec:dynamics}

\subsection{Expectations for FRIIs}
\label{sec:FRIIdynamics}

The energy and momentum fluxes in FRII jets are
expected to be sufficient to drive a bow shock at supersonic
speed into the ambient medium \citep[e.g.,][]{leahy}. Ambient gas crossing
the bow shock will be heated.  For a shock advance speed relative to the
speed of light of $v_{\rm adv}/c$, the Mach number, ${\cal M}$, in monatomic gas of
normal cosmic abundances with thermal energy $kT$ in units of keV, is given by

\begin{equation}
{\cal M} \sim 580 (v_{\rm adv}/c) (kT)^{1/2}.
\end{equation}

\noindent
For a non-relativistic equation of state ($\gamma = 5/3$),
the jump conditions for a non-radiating shock \citep[e.g.,][]{spitzer}
find that
pressure, density, and temperature ratios between gas that
has crossed the shock and the ambient medium are

 \begin{equation}
P_2/P_1 = (5 {\cal M}^2 -1)/4
\label{eq:rhpress}
\end{equation}

\begin{equation}
\rho_2/\rho_1 = 4 {\cal M}^2/ ({\cal M}^2 +3)
\label{eq:rhdens}
\end{equation}

\begin{equation}
T_2/T_1 = (5 {\cal M}^2 -1)({\cal M}^2 +3)/16 {\cal M}^2
\label{eq:rhtemp}
\end{equation}

\noindent
where subscripts 2 and 1 refer to post-shock and pre-shock conditions, respectively.
For high advance speed and large Mach number the density contrast
reaches a factor of four, resulting in enhanced X-ray
emissivity from shocked gas.  The visibility in observations will
depend on the relative volumes of shocked and unshocked
gas along given lines of sight.

Complications apply in reality.
Firstly, there is observational evidence that in
supernova remnants the post-shock electrons are cooler than the ions
\citep[e.g.,][]{hwang-snr, rakowski-snr}.  Secondly, a bow shock around a lobe
is oblique away from its head, with a
consequent change in the jump conditions and the emissivity contrast
\cite{williams}.   The closer a structure is to
a spherical expansion, the more normal the shock will be everywhere
and the better the applicability of the above equations.

\rosat\ data revealed the presence of X-ray cavities coincident with
the inner parts of the radio lobes of Cygnus~A, and these were
interpreted as due to the contrast between undisturbed ambient gas and
gas around the lobes that had been heated in the past but has now
expanded and cooled to a low emissivity \cite{carilli-cyga}, although
the parameters of the shock are not effectively constrained by the
data.  More recent {\it Chandra\/} observations of Cygnus~A find gas
at the sides of the lobes to have $kT \sim 6$~keV, slightly hotter
than the value of 5~keV from ambient medium at the same cluster
radius,
but the gas may have cooled after bow-shock heating,
and again the data do not usefully constrain model parameters
\cite{smith-cygacluster}.  Evidence of strong shock heating around more
distant FRII radio galaxies has yet to be seen.

CSS and GPS sources have been examined for evidence of shock heating.
These are good places to look as the radio sources are generally
considered to be in an early stage of expansion and they are
overpressured with respect to even a cluster ambient medium
\citep[e.g.,][]{siem-3c186}.  The disadvantage is that source sizes
are small so that even \chandra\ will have difficulty in separating
emission from the nuclei, radio structures, and ambient medium from
that of any shocked gas.  The best evidence for detection of shocked
gas thus arises from deep XMM-Newton spectroscopy, and in particular
that of the CSS source 3C~303.1 \cite{odea-303.1}.  The X-ray spectrum
contains soft emission (associated with the ambient galaxy atmosphere)
and a hard component.  Since nuclear emission is undetected in the
radio, it is reasonable to associate the hard emission with shocked
gas, and a model can be constructed \cite{odea-303.1} that has an
expansion velocity consistent with cooling-time arguments for optical
emission-line gas \cite{devries}.

\subsection{Dynamics of FRIs in clusters}
\label{sec:FRIdynamicscluster}

Low-power sources are closer and more amenable to detailed study,
since the various components of X-ray emission are
more easily separated.
The medium plays an important r\^ole in the deceleration of the jets,
which share momentum and energy with entrained material
(\S\ref{sec:jetcollimate}).  

The \einstein\ and \rosat\ missions found evidence that the radio
lobes of NGC~1275 have pushed Perseus-cluster gas aside
\cite[e.g.,][]{boer-n1275rosat}, and now many clusters and groups are
found to harbour gas cavities containing radio plasma that originates
from active galaxies \citep[e.g.,][]{birzan}.  Rather than expansion
at high Mach number, the displacement of the gas appears normally to
create low-density, rising bubbles in rough pressure balance with the
surrounding medium \citep[e.g.,][]{churazov-m87}.  NGC~1275, M~87, and
Hydra~A are showcase examples with deep \chandra\ exposures and
complex bubble and cavity systems \cite{fabian, forman, wise}. 
Radio bubbles in clusters are sufficiently common that they are
an important heat source today, with enough power to
balance the radiative cooling of dense gas in clusters
\citep[e.g.,][]{dunn, rafferty}, although the total energies and
lifetimes of individual bubbles are considerably uncertain.  An issue
of particular interest that follows from this is the potential for the
associated heating and cooling to forge the link between black-hole
and galaxy growth.  A recent review is available
\cite{mcnamara-nulsen}, and so the topic is not dealt in depth here.

It is noteworthy that the luminosity function of radio sources places
the energetically dominant population to be roughly at the FRI/FRII
boundary \cite[e.g.,][]{ledlow-owen}, rather than within the more
numerous but lower power population of FRIs studied in nearby clusters
(although there are claims that total jet power scales slightly less than
linearly with radio power \citep[e.g.,][]{willott, birzan2}).  It thus
remains possible that the rather gentle heating around currently
studied sources does not provide us with the complete picture, and
violent shock heating around more powerful sources is
energetically important but currently eluding detection.

\subsection{Centaurus~A}
\label{sec:dynamicscenA}

\begin{figure}[t]
\centering
\includegraphics[height=8.8cm]{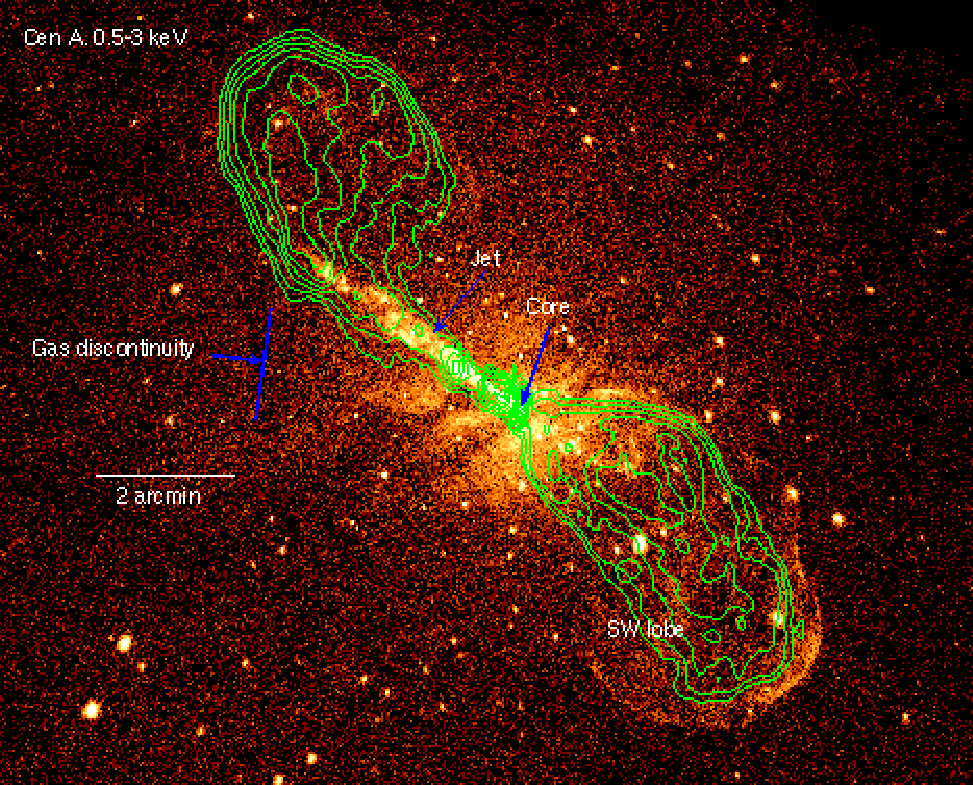}
\caption{Radio contours on a deep \chandra\ image of Cen~A, showing
the core and NE jet crossed
by absorption stripes corresponding to NGC~5128's dust lanes, 
the SW lobe, structures associated with the NE lobe, the
position of a merger-related gas discontinuity that shows up better at
lower energies, and many XRBs in NGC~5128
\protect\citep[see][]{hard-cenarecent, jordan-cenarecent,
worr-cenarecent, kraft-cenarecent, siv-cenarecent}.}
\label{fig:cenasource}      
\end{figure}

The best example of supersonic expansion is not in an FRII radio
source but associated with the inner southwest radio lobe of Cen~A
\citep[][and see Fig.~\ref{fig:cenasource} for a more recent, deeper,
\chandra\ image]{kraft-cenalobe}.  Cen~A is our nearest radio galaxy,
where 1~arcmin corresponds to $\sim 1.1$~kpc.  The full extent of
Cen~A's radio emission covers several degrees on the sky
\cite{junkes-cena}.  Within this lies a sub-galaxy-sized double-lobed
inner structure \cite{burns-cena} with a predominantly one-sided jet
to the NE and weak counter-jet knots to the SW \cite{hard-cena} that are
embedded in a radio lobe with pressure at least ten times larger than
that of the ambient ISM \cite{kraft-cenalobe}.  The lobe should be
expanding and be surrounded by a shock.  The associated structure is
exquisitely seen in Figure~\ref{fig:cenasource}.  Although the capped
SW lobe is around the weak counterjet, so it is not evident that the lobe
is being thrust forward supersonically with respect to the external
interstellar medium (ISM) by the momentum flux of an active jet, the
high internal pressure in the radio lobe ensures its strong expansion.

The density contrast between post-shock and pre-shock gas in Cen~A inferred by
\cite{kraft-cenalobe} was larger than four, which is not allowed by
Equation~\ref{eq:rhdens}, and so straightforward modelling was not possible.
New modelling is underway using results from the new deep observation.
However, the conclusion that the lobe's kinetic energy exceeds its
thermal energy, and the thermal energy of the ISM within 15 kpc of the
centre of the galaxy, is unlikely to change.  As the shell dissipates,
most of the kinetic energy should ultimately be converted into heat
and this will have a major effect on Cen~A's ISM, providing
distributed heating.

There is much still to be learned about how gas is displaced by radio
structures, and the processes of heat transfer.  A new view will be
possible with the high-resolution spectroscopic capabilities of the
International X-ray Observatory currently under study by ESA and NASA.
This will provide the vital ingredient of useful velocity data, giving
a handle also on such issues as turbulence and
non-perpendicular velocities at shocks.

\subsection{The effect of galaxy mergers}
\label{sec:mergers}

It is important to understand what triggers radio activity and what
causes it to cease, particularly since radio sources are now
recognized as an important heat source for large-scale structure
(\S~\ref{sec:FRIdynamicscluster}).
It has long been recognized that mergers may be important in
{\it triggering\/} radio activity, and this is consistent with the preference
for low-power radio galaxies to reside in clusters and rich groups.
For example, NGC~1275 and M~87 (\S~\ref{sec:FRIdynamicscluster}) are
the dominant galaxies of the Perseus and Virgo clusters, respectively.
Cen~A (\S~\ref{sec:dynamicscenA}) is hosted by NGC~5128 which in turn
hosts an inner warped disk
suspected to be the merger remnant of a small gas-rich spiral galaxy
\citep[e.g.,][]{quillen}.

Mergers leave an imprint on the temperature, density, and metallicity
structures of the gas.  Due to good linear resolution it is again
Cen~A that shows such effects particularly well, with clear
indications that even the hot X-ray-emitting gas is poorly mixed.  The
merger appears to be having an important influence on the evolution of
the northeast radio jet and inner lobe \cite{kraft-cenarecent}.

\begin{figure}[t]
\centering
\includegraphics[height=6.8cm]{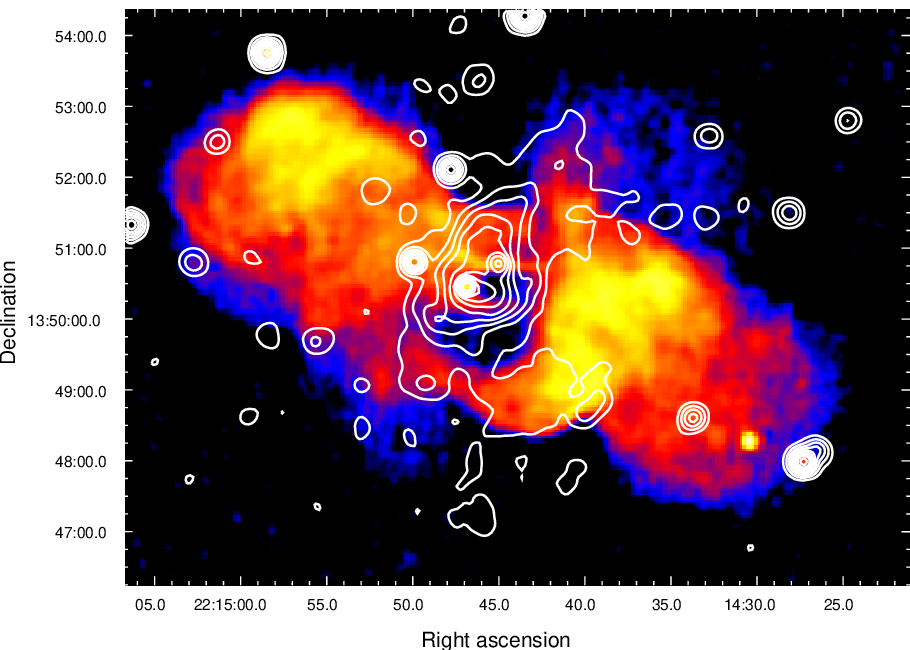}
\caption{ \chandra\ contours (logarithmic spacing) on a color
radio image of 3C~442A, taken from \protect\cite{worr-dead}.
Bright X-ray emission from the merger atmospheres of
NGC\,7236 and NGC\,7237 fills the gap between the radio lobes
which are no longer fuelled by an active jet.
}
\label{fig:3c442a}      
\end{figure}

In the more extreme case of 3C~442A (Figure~\ref{fig:3c442a})
there is evidence that a merger
may have smothered a previously active jet, leaving a large volume of
decaying radio plasma, while at the same time re-starting jet activity
in the nucleus of one of the galaxies 
\cite{worr-dead}.
Here the merger gas has sufficiently high pressure for the radio lobes
to be riding on the pressure front of the merger gas that is sweeping
them apart.  The energy in the
merger gas will eventually be dissipated in the outer regions of the
group atmosphere --- an additional source of heating to that arising
from both the old and new merger-induced radio activity.  The radio
spectrum from the old decaying radio lobes is flatter where they are
being compressed by the expanding merger gas, suggesting that energy
from the gas has a second effect, in re-exciting relativistic electrons
through compression and adiabatic heating \cite{worr-dead}.
While it is undoubtedly true that mergers produce messy substructures,
the
example of 3C~442A suggests that there is some prospect that the
switching on and off of radio activity
by mergers can be timed (albeit roughly) using the morphology of the
stellar component of the galaxies and spectral changes in the radio
plasma, and that this can be combined with the measured energy content of the gas
and radio plasma to trace the history of radio outbursts and their
effectiveness in heating gas.

\section{Is a jet's fate determined by the central engine?}
\label{sec:engine}

\subsection{An evolutionary cycle?}
\label{sec:evolution}

The Ledlow-Owen relation (\S \ref{sec:classes}) showed that a galaxy
of a given optical luminosity can host either an FRI or FRII radio
source. This resulted in renewed speculation in the 1990s that there
may be an evolution between FRII and FRI activity controlled by
external influences.  Such speculation was supported by evidence of
FRIIs associated with galaxy mergers (distorted isophotes and higher
amounts of high-excitation ionized gas) and FRIs associated with
galaxies in more relaxed dynamical states \cite{bick-evol}.
Evolutionary ideas have also arisen from the so called `fundamental
plane' that places AGN on an extension of the relationship between
inner jet radio power, X-ray luminosity and black-hole mass found for
X-ray binaries (XRBs) \cite{merloni, falcke}.  It has been suggested
that the changes in X-ray spectrum and jet luminosity that accompany
changes in accretion characteristics in an XRB could apply to AGN,
such that an individual object may go though transitions between an
FRI and FRII, and indeed to becoming radio quiet
\citep[e.g.,][]{kording}.

Observationally, kpc-scale jets accompany AGN with accretion flows
that in the extreme are either geometrically-thick and radiatively
inefficient or geometrically-thin and radiatively efficient, with the
latter accompanied by high-excitation optical emission lines.  It is
possible that an AGN changes over the lifetime of a radio source, such
that the observed kpc-scale radio structures are the result of
ejection from an AGN evolving through different states. 
Some sort of intermittency of the
central engine over timescales
of $\sim 10^4 - 10^6$ years (shorter than the lifetime of radio
sources, \S\ref{sec:dutycycles}) gains support from observational and
theoretical considerations \citep[e.g.,][]{reybeg, schoenmakers, janiuk, stawarz}.
Multiple changes to the central structure over the
lifetime of the radio source
would be required to reconcile the claim that a geometrically-thick
flow is needed to sustain a significant jet (with the most powerful
requiring a spinning black hole) \cite{meier} with the observation
that many AGN with powerful jets currently show geometrically-thin disks and
high-excitation emission lines (see below).  

Closer examination is needed of the extent to which the observed
powers and structures of jets relate either to the accretion processes
or to large-scale environmental effects.  Both appear to play a r\^ole.

\subsection{The r\^ole played by accretion processes}
\label{sec:accretionmode}

Broadly, powerful jets of FRII structure are associated with AGN
showing high-excitation optical emission lines, while lower-power
jets, normally but not always of FRI structure, are associated with
AGN showing low-excitation lines.  This suggests that the central
engine has at least some influence on the power and large-scale
structure of the jets \citep[e.g.,][]{baum}.

A correlation between the core radio emission and low-energy ($\sim 1$
to 2 keV) nuclear X-ray output of radio galaxies has been known since
the \einstein\ and \rosat\ missions, and has been used to argue that
the soft X-rays arise from pc-scale jets \cite{fabbiano-3c,
worr-rosfr1, canosa, hard-worr}.  An optical core is often seen with
\hst, and is interpreted as synchrotron emission from a similar
small-scale emitting region \cite{chi-optcores, hard-worroptcore,
capetti-b2optcores, chi-optcoresFRII, verdoes-kleijn}.  Such pc-scales
jets protrude from any gas and dust torus invoked by AGN unified
models, and so this component should not be greatly affected by
absorption, although relativistic effects will cause jet orientation
to affect the level of X-ray flux observed.

Since jet emission dominates at low
X-ray energies, it has been important to obtain sensitive spectral
measurements that extend to the higher X-ray energies accessible to
\chandra\ and \xmm\ in order to probe the region closer to the SMBH and
representative of the bolometric power of the central engine.  At
these energies
any strong emission from the AGN should dominate jet emission even if
it is largely absorbed at lower energies by a gas torus.  Results find
a number of radio galaxies showing clear evidence of a hard continuum,
sometimes accompanied by Fe-line emission, and presumed to be emission
associated with an accretion-disk corona \citep[e.g.,][]{ueno, young,
glioz-n4261core}.  Both the jet and central-engine X-ray components
can sometimes be distinguished in the same spectrum
\citep[e.g.,][]{croston2, evans-cena, zezas-4261}.

The hard component is more often detected in FRIIs than in FRIs.  Of
course, greater absorption from a torus could potentially combine with
lower X-ray luminosity in causing the non-detection of the second
component in most FRIs, and so particular reliability can be placed on
the results of a study of nearby ($z < 0.1$) radio galaxies that has
allowed for absorption in placing upper limits on the luminosity of
undetected nuclear components \cite{evans-cores}.  The radiative
efficiency of the central engine was then found by correcting the
X-ray luminosity to a bolometric luminosity and combining it with the
inferred SMBH mass.  In powerful FRIIs, radiatively-efficient
accretion associated with a thin disk surrounded by an obscuring torus
is normally inferred.  FRII radio galaxies at $z \sim 0.5$ also show
an absorbed X-ray component \cite{belsole-cores}.  In contrast, in $z
< 0.1$ FRIs, all the nuclear X-ray emission can normally be
interpreted as jet related, and usually only upper limits are found
for accretion-related emission \cite{evans-cores}.  Any X-ray
luminosity associated with a non-jet central-engine component in
low-power sources is normally sufficiently low to support earlier
speculations based on the Ledlow-Owen relation that the physical
difference between the two types of radio source arises from the
different nature of their accretion disks and efficiency of accretion
\cite{ghisellini-celotti}.  Further support for these ideas comes from
{\it Spitzer\/} results for the $z < 0.1$ sample \cite{birk-spitzer}
that show an additional component of hot dust only in FRIIs.

While results at first-look appear quite convincing of a connection
between large-scale radio power and the structure of the central
engine, there are sources which defy the trend.  Both Cen A and NGC
4261 have large-scale FRI structures, and yet contain absorbed, hard,
luminous X-ray components characteristic of the coronae of thin
accretion disks seen through an obscuring torus \cite{evans-cena,
zezas-4261}.  This might suggest that something relatively recent
(perhaps the galaxy merger in the case of Cen A \cite{evans-cores})
has provided additional material for accretion and affected the
central engine in a way that has yet to be reflected in the power and
structure of the large-scale radio emission. The difficulty is that
merger and source-development time scales are expected to be
comparable.  A further complication is the tendency for any X-ray
accretion-related components in FRII low-excitation radio galaxies to
be less luminous than those seen in a typical FRII high-excitation
radio galaxy \cite{hardevc}, as was known for the optical continuum
\cite{chi-optcoresFRII, varano}.  This means that not all FRIIs have
equivalent central engines.  However, it is is hard to treat as a
coincidence the tendency for the most powerful FRIIs with the least
evidence for external disruption to arise from AGN showing
high-excitation optical emission lines and evidence for thin accretion
disks.

In the normally inferred absence of thin radiatively-efficient
accretion disks in FRIs, it has been argued in several cases that sufficient
X-ray-emitting hot gas is present in their galaxies and clusters to
produce the required jet power through a geometrically-thick Bondi
accretion flow \citep[e.g.,][]{dimatteo, allen}. Here the jet power is
inferred from the energy required to excavate the cavities observed in
the X-ray-emitting gas, i.e., a more direct method than scaling from
radio power \citep[e.g.,][]{willott} as is normal in the absence of
other information.  Recent work confirms that the most luminous FRIIs
also tend to lie in luminous X-ray clusters \cite{belsole-env}, and it
is reasonable to assume that they experience similar or greater
supplies of galaxy and cluster hot gas.  However jet powers are also
higher (how much so rests on uncertainties in speed and composition),
consistent with requiring an extra energy source in the form of stars
and gas clouds fuelling a thin accretion disk.  A major outstanding
problem is a full understanding of the mechanisms which convert gas
infall into two different accretion structures.  Jets are expected to
be more strongly coupled to the structure of the host stellar system,
and hence to play a more major r\^ole in feedback, if the accreting
gas originates predominantly from the reservoir contained in the
potential well of the system as a whole, whether it be hot
\citep[e.g.,][]{allen} or cold \citep[e.g.,][]{rawlingsjarvis} in
origin.

\subsection{The r\^ole of the environment}
\label{sec:environment}

Assuming that jets are genuinely symmetric at production, the
environment appears to be, at a minimum, a strong secondary factor
(with jet power being the likely primary influence) in shaping
large-scale jet structure.  For example, some radio sources show what
appears to be FRI morphology on one side and FRII on the other, and
this has been used to argue for different
environmental effects on the two sides \cite{gopal-wiita}.

VLBI proper-motion studies find few, if any, differences in the
speed or morphology of FRI and FRII radio jets in their initial stages
of development from the central engine \cite{pearson, giovannini}.
However, the radiative powers are higher in FRIIs, but not in linear
proportion to their total radio powers \citep[e.g.,][]{giovanninietal,
giovanninietal2}, suggesting that on the small
scale a radio source has knowledge of how it will evolve.
Particularly compelling evidence that the environment does have some influence is the
recent discovery that quasars, traditionally the hosts only of FRII
structures, can host FRI radio structures, with evidence that denser,
more clumpy, environments at higher redshift are allowing this to
occur \cite{heywood}.  The r\^ole of the X-ray-emitting environment in
decelerating FRI jets was discussed in \S\ref{sec:jetcollimate}.

\subsection{Information from beamed sources}
\label{sec:beamed}

The beamed counterparts of radio galaxies (quasars and BL Lac objects)
do not allow the accretion structures to be probed in the X-ray, since
the beamed jet emission swamps all other nuclear components; indeed it
is sometimes dominant up to the TeV band.
Multi-wavelength spectral energy distributions and variability time
scales are used to probe the beaming parameters and the physical
properties of the emitting regions \citep[e.g.,][]{ghisellini-model,
krawczynski-mrk421tevvar, tagliaferri-var}.  Correlated flares are
sometimes measured across wavebands, giving support to the presence of
a dominant spatial region of emission \citep[e.g.,][]{urry-2155var,
takahashi-mrk421var}, but otherwise uncertainties of size scales,
geometries, and parameters for the competing processes of energy loss
and acceleration often force the adoption of oversimplified or
poorly-constrained models for individual jets.  Much is published on
the topic, and a review is beyond the scope of this work.
Substantial progress
in understanding is anticipated from multiwavelength programmes
associated with the Fermi Gamma-ray Space Telescope.

VLBI radio-polarization studies have found systematic differences
between powerful quasars (beamed FRIIs) and BL Lac objects (beamed FRIs) in
core polarizations, the orientations of the magnetic fields in the
inner jets, and in jet length, although it is difficult to separate
intrinsic differences from the possible influence of the parsec-scale
environment, such as the density and magnetic field contained in line-emitting
gas \cite{cawthorne}.

\section{Summary and concluding remarks}
\label{sec:conclusions}

The last decade has seen massive progress in our understanding
of the X-ray properties of extragalactic radio jets and their
environments.  \chandra's sub-arcsec spatial resolution has been of
paramount importance in measuring resolved X-ray emission from
kpc-scale jet structures, and in extending studies of X-ray nuclei to
sources other than beamed quasars and BL~Lac objects by separating
the emission of weaker nuclei from that of the jets and
X-ray emitting environments.  

The assumption that radio structures roughly lie in a state of minimum
energy between their relativistic particles and magnetic fields is
broadly verified in a few tens of sources through combining X-ray
inverse Compton with radio synchrotron data (\S\ref{sec:minenergytest}).
This is the assumption commonly adopted in the absence of other
information, and so its verification is reassuring, although much
sub-structure is likely to occur and there is no reason to expect
minimum energy to hold in dynamical structures.

The increase in numbers of known resolved kpc-scale X-ray jets has
been remarkable, from a handful to the several tens of sources that
\chandra\ has mapped in detail.  There are grounds to believe that
there are X-rays from synchrotron radiation in sources both of FRI and
FRII types (\S\ref{sec:Xsyn}), requiring in-situ particle acceleration
to TeV energies.  The steepening in spectral slope which most commonly
occurs at infra-red energies may be related more to acceleration
processes than energy losses, but more multiwavelength observational
work is required to characterize the acceleration sites and support a
theoretical understanding.  The fact that X-ray synchrotron emission
with an X-ray to radio flux-density ratio, $S_{\rm 1~keV}/S_{\rm 5~GHz}$,
between about $10^{-8}$ and $10^{-7}$ is so common in jets where the
bulk flow is inferred to be relativistic implies that there will be
many more X-ray jet detections with current instrumentation in
sufficient exposure time.

The dominant X-ray emission mechanism in resolved quasar jets
remains uncertain, but it is likely that beamed emission from
scattering of CMB photons is dominant in jets at small angles to the
line of sight.  This requires that highly relativistic
bulk flows exist far from the cores, contradicting
earlier radio studies but possibly understandable in the context of
transverse velocity profiles.
The knotty appearance of these jets is then possibly a result of variable
output from the nuclei.
Much of the knotty X-ray appearance of FRI jets, on the other hand, likely
arises from spatial variations in the strength of particle
acceleration (\S\ref{sec:CenAn315}).

Jet theory has had some
pleasing successes, such as the agreement between X-ray pressure
profiles and predictions from hydrodynamical models for low-power
jets in the regions where they are believed to be slowed by entrainment of the
external medium or stellar mass loss (\S\ref{sec:jetcollimate}).

We are still largely ignorant of jet composition, and this is a
difficult problem to solve since jet dynamics are governed by the energy
of the constituent particles rather than their mass.  There is generally growing
support for a strong presence of relativistic protons
(\S\ref{sec:jetcomposition}).

The observation of bubbles and cavities in cluster gas produced
dynamically by radio structures has renewed interest in the mechanisms
by which active galaxies introduce heat into gaseous atmospheres.  A few
nearby bright systems have been the subject of intense study with
\chandra\ (\S\ref{sec:FRIdynamicscluster}).  Although the way in which
energy is deposited on the large scale is still far from clear,
information on morphology and temperature has been used to infer the
underlying energetics of the structures.

An area where work is still in its infancy is that of understanding
the triggering of radio sources, and the possible r\^ole played here
by galaxy and cluster mergers in promoting or inhibiting radio-source
development (\S\ref{sec:mergers}).  The emerging picture shows that
very different accretion structures can host radio jets, with a
tendency for quasar-type nuclei to be associated with more powerful
jets.  How jets are powered by these different accretion structures
and gas infall, and the duration of a given mode relative to typical
lifetimes of radio sources, remain to be better understood.

The future is bright.  \chandra\ and \xmm\ are now mature
observatories.  Operational experience is enabling both
more ambitious and more speculative programs to be undertaken.  For
example,  \chandra\ is
completing sensitive exposures of all 3CRR
radio sources within a redshift of 0.1, and a
large shallow survey of quasar jets to study the X-ray-emission
mechanism in a statistical sense and seek out more sources for deep,
detailed study.  Observations of a somewhat more speculative
nature are also being made, such as observing radio sources of
different inferred ages, and
studying how galaxy and cluster mergers are impacting the radio-source
structures and their influence on the surrounding atmospheres.  These
are just examples.  At the same time, 
\suzaku\ is making spectral measurements of
active-galaxy nuclei, and testing the spin characteristics of black holes
hosting radio sources through searching for relativistic broadening in Fe
lines.   We can expect fantastic results from continuing X-ray work, and many
surprises.

New facilities coming on line will enrich the X-ray results. 
\spitzer\ has measured dust, stars, and non-thermal cores in
the centres of radio galaxies, placing constraints on the central
structures.  It has also detected a number of kpc-scale jets, helping
to tie down the all-important breaks in the spectral distributions of
the synchrotron radiation that are likely to be connected to the
process of particle acceleration.  \herschel\ will continue such work.

The characteristics of the non-thermal emission at energies higher
than the X-ray provide a sensitive test of emission mechanisms and a
probe of jet composition.  The Fermi Gamma-ray Space Telescope
is providing such data, particularly for the
embedded small-scale jets of highly-beamed quasars and BL~Lac objects,
as are ground-based Cerenkov telescopes sensitive to TeV emission.

\alma\ will probe the cool component of gas in active galaxies, and
provide information on one possible component of accretion power.
Radio measurements with \emerlin\ and \evla\ will probe spatial scales
intermediate between pc and kpc, important in the
launching and collimation of jets.  They will also provide improved
information on transverse jet structure.

Extending polarimetry to the X-ray, as is under study in the
community, will provide key tests of jet emission and acceleration
mechanisms, just as such work with \hst\ is starting to do in the
optical.  Most importantly, a future X-ray observatory that has the
sensitivity and spectral resolution to probe gas dynamics associated
with radio sources is crucial for confirming and extending source
modelling that is currently in its infancy.  Such capabilities will
come with the launch of a new facility such as the International X-ray
Observatory currently under study by ESA and NASA.

\begin{acknowledgements}
I am grateful to Mark Birkinshaw for his essential contributions to
our collaborative work on radio sources since the early 1990s, when we
first observed radio galaxies with \rosat.  More recently, colleagues
and students too numerous to list have energetically worked on jet
data, and helped feed my enthusiasm for the subject.  I particularly
thank all involved in making \chandra\ such a great success,
permitting resolved X-ray emission to be seen so clearly in so many
active-galaxy jets. The outline for this review has evolved from talks
I gave at `6 Years with \chandra', Cambridge MA, November 2005, and
`Observations from High Energy Astrophysics Satellites' at the Marcel
Grossmann meeting, Berlin, July 2006, and I am grateful to the
organizers of those meetings for their invitations.  I thank Raffaella
Morganti, Thierry Courvoisier and Mark Birkinshaw for suggestions that
have improved the manuscript.  The NASA Astrophysics Data System has
greatly assisted me in constructing the bibliography for this review.
\end{acknowledgements}


\end{document}